 \documentclass[pre,amsmath,amssymb,twocolumn, showpacs, superscriptaddress,10pt]{revtex4-1}

\usepackage{amsmath}
\usepackage{hyperref}
\usepackage{graphicx}
\usepackage{amsfonts}
\usepackage{amsthm}
\usepackage{cases}
\usepackage{bm}

\usepackage{fixmath}
\newcommand{\vect}[1]{\mathbold {#1}} 

\usepackage{enumitem} 

\usepackage[normalem]{ulem}

\usepackage{color}
\definecolor{Blue}{rgb}{0.00, 0.00, 0.80}
\definecolor{Red}{rgb}{0.80, 0.00, 0.00}
\definecolor{Green}{rgb}{0.00, 0.50, 0.00}

\hypersetup{
    colorlinks=true,       
    linkcolor=red,          
    citecolor=blue,        
    filecolor=magenta,      
    urlcolor=cyan           
}

\newcommand{\antiquad}{\!\!\!\!\!\!\!\!}
\newcommand{\nn}{\nonumber}
\newcommand{\be}{\begin{equation}}
\newcommand{\ee}{\end{equation}}
\newcommand{\bea}{\begin{eqnarray}}
\newcommand{\eea}{\end{eqnarray}}
\def\la{\langle}
\def\ra{\rangle}

\newcommand{\beq}{\begin{equation}}
\newcommand{\eeq}{\end{equation}}
\newcommand{\beqn}{\begin{eqnarray}}
\newcommand{\eeqn}{\end{eqnarray}}

\begin{document}

\title{Exact position distribution of a harmonically-confined run-and-tumble particle in two dimensions}

\author{Naftali R. Smith}
\email{naftalismith@gmail.com}
\affiliation{Department of Solar Energy and Environmental Physics, Blaustein Institutes for Desert Research, Ben-Gurion University of the Negev, Sede Boqer Campus, 8499000, Israel}
\author{Pierre Le Doussal}
\email{pierre.ledoussal@phys.ens.fr}
\affiliation{Laboratoire de Physique de l’Ecole Normale Sup\'erieure, CNRS, ENS \& Universit\'e PSL, Sorbonne Universit\'e,
Universit\'e de Paris, 75005 Paris, France}
\author{Satya N. \surname{Majumdar}}
\email{satya.majumdar@universite-paris-saclay.fr}
\affiliation{Universit{\'e} Paris-Saclay, CNRS, LPTMS, 91405, Orsay, France}
\author{Gr\'egory \surname{Schehr}}
\email{schehr@lpthe.jussieu.fr}
\affiliation{Sorbonne Universit{\'e}, Laboratoire de Physique Th{\'e}orique et Hautes Energies, CNRS UMR 7589, 4 Place Jussieu, 75252 Paris Cedex 05, France}
\date{\today}

\begin{abstract}
We consider an overdamped run-and-tumble particle in two dimensions, with self propulsion in an orientation that stochastically rotates by 90 degrees at a constant rate, clockwise or counter-clockwise with equal probabilities. In addition, the particle is confined by an external harmonic potential of stiffness $\mu$, and possibly diffuses. We find the exact time-dependent distribution $P\left(x,y,t\right)$ of the particle's position, and in particular, the steady-state distribution $P_{\text{st}}\left(x,y\right)$ that is reached in the long-time limit. We also find $P\left(x,y,t\right)$ for a ``free'' particle, $\mu=0$.
We achieve this by showing that, under a proper change of coordinates, the problem decomposes into two statistically-independent one-dimensional problems, whose exact solution has recently been obtained.
We then extend these results in several directions, to two such run-and-tumble particles with a harmonic interaction, to analogous systems of dimension three or higher, 
and by allowing stochastic resetting.
\end{abstract}



\maketitle

\section{Introduction}

Active particles consume energy from their environment and use it in order to generate dissipated directed motion  \cite{Romanczuk,soft,BechingerRev, FodorEtAl15, Fodor16, Needleman17, Ramaswamy2017,Marchetti2017,Schweitzer, FJC22}. Examples of active matter are ubiquitous in nature, including many biological systems of living cells and/or bacteria  \cite{Berg2004,Mizuno07, Wilhelm08, Cates2012, Ahmed15, Breoni22}, flocks of birds \cite{flocking1, flocking2} fish schools \cite{Vicsek,fish}, and also physical systems such as granular matter \cite{gran1,gran2, Bruot11}.
Activity breaks time-reversal symmetry, and therefore, drives the system out of thermal equilibrium.
Active matter has attracted much interest over recent years, which led to the discovery of several remarkable collective behaviors that are very different to those observed in systems in thermal equilibrium. These behaviors include motility induced phase separation \cite{separation1, separation2, separation3}, clustering \cite{cluster1,cluster2,evans}, and the absence of an equation of state relating pressure to the system's bulk properties  \cite{Kardar2015}.

In fact, even at the level of a single particle, active particles display some nontrivial features that are not observed in their passive counterparts.
In particular, active particles that are affected by an external potential have been recently studied, both theoretically \cite{Franosch2016,Das2018,Caprini2019,Sevilla2019} and experimentally \cite{Hagen2014,Takatori,Deblais2018,Dauchot2019}.
It was shown that such a particle can reach a non-Boltzmann steady state, and/or cluster near the boundaries of a spatial region in which it is confined \cite{Solon2015,Potosky2012,ABP2019,Dhar_2019,Malakar20}, and that it develops a nonzero drift velocity even if the external potential is periodic \cite{led20}. First-passage and relaxation properties were also studied \cite{MalakarEtAl18,ABP2018, Singh2019}.

In order to make progress analytically, it is usual to focus the study on simple theoretical models. One such model of active particles, that has been extensively studied, is the model of the run-and-tumble particle (RTP). This model describes an overdamped particle whose speed $v_0$ is constant, while the orientation of its velocity changes in time randomly via sudden jumps (or `tumbles').
In one spatial dimension (1D), this model becomes especially simple: the only possible velocity orientations are $\sigma = \pm 1$, i.e., the particle's velocity can be $\pm v_0$. At a constant rate $\gamma$, the orientation flips $\sigma \to -\sigma$.
One can optionally take into account an external potential $U(x)$ too. The position $x(t)$ of this RTP obeys the Langevin equation
\be
\dot x = f(x) + v_0 \sigma(t) \, . \label{eq:RTP_2st}
\ee
Here $f(x)= - U'(x)$ is the deterministic force exerted on the particle due to the external potential $U(x)$, while the orientation $\sigma$ plays the role of a telegraphic (dichotomous) noise. The statistical properties of $\sigma$ lead to a breaking of time reversal symmetry. In contrast to the white (Gaussian) noise in equilibrium systems,  $\sigma(t)$ is a colored noise; its autocorrelation function is $\la \sigma(t) \sigma(t') \ra = e^{-2 \gamma |t-t'|}$ (angular brackets denote ensemble averaging), describing exponential decay with a typical timescale of  $\tau = (2\gamma)^{-1}$.
Many properties of the 1D RTP can be found exactly, as we recall shortly. However, despite its apparent simplicity, the model displays many nontrivial features, e.g., a steady-state distribution that is non-Boltzmann \cite{Dhar_2019, MalakarEtAl18}. 

One of the most fundamental quantities to study is the (time-dependent) position distribution $P(x,t)$ of the particle, given that it is initially at the origin $x(t=0) = 0$. Let us assume that the initial orientation is randomly selected from the two possible values, $\sigma(0) = \pm 1$, each with equal probability $1/2$. 
For an RTP in 1D that is `free', i.e., in the absence of an external potential (so $f(x) = 0$),
$P(x,t)$ is known exactly, but is nevertheless highly nontrivial \cite{wei02,HV_2010,ODA_1988,MADB_2012,MalakarEtAl18,EM_2018,Dhar_2019,SBS20, Dean21}.
 The support of the distribution is is the
interval $x\in [-v_0\, t, v_0\, t]$ and for $|x|\le v_0\, t$ it is given by
\bea
\label{p1xt.1}
P_{\text{free}}\left(x,t\right)&=&\frac{{\rm e}^{-\gamma t}}{2}\biggl\{ \delta\left(x-v_{0}t\right)+\delta\left(x+v_{0}t\right) \nn\\
&+&\left.\frac{\gamma}{2v_{0}}\left[I_{0}(\rho)+\frac{\gamma I_{1}(\rho)}{\rho}\right]\theta\left(v_{0}t-|x|\right)\right\} \,,
\eea
where
\be
\rho = \sqrt{v_0^2 t^2 - x^2}\,\frac{\gamma}{v_0}
\ee
and $I_0(\rho)$ and $I_1(\rho)$ are modified Bessel functions of the
first kind. 
The $\delta$ functions at the edges of the support $x=\pm v_0\, t$ correspond to
the cases where $\sigma(0) = \pm 1$ (respectively) and the noise $\sigma(t)$ does not change its value up to time $t$.
At long times, the central part of the distribution approaches a Gaussian form, as one would expect since the free RTP reduces, at late times, to ordinary Brownian motion.

The presence of a confining potential complicates the theoretical analysis considerably. Nevertheless,  the (nonequilibrium) steady-state distribution of the RTP's position  is known exactly for an arbitrary confining potential $U(x)$. It is given, up to a normalization constant, by
\be
\label{eq:Pst_2st}
P_{\textrm{st}}(x)\propto\frac{1}{v_{0}^{2}-f^{2}(x)}\exp\left[2\gamma\int_{0}^{x}dy\frac{f(y)}{v_{0}^{2}-f^{2}(y)}\right]
\ee
The result \eqref{eq:Pst_2st} has been known for decades, obtained originally in the context of quantum optics \cite{q-optics1,q-optics2,q-optics3,q-optics4} and later reproduced in the study of colored noise on dynamical systems \cite{VBH84, colored} and of active matter \cite{Kardar2015, Dhar_2019}.
In the diffusive limit, when $v_0 \to \infty$, $\gamma \to \infty$ but keeping the ratio $v_0^2/2\gamma = D$ fixed, the dynamics converge to the overdamped dynamics of a particle of diffusivity $D$ in a trapping potential $U(x)$. Indeed, one finds that in this limit the distribution \eqref{eq:Pst_2st} reduces to a Boltzmann distribution $P_{\textrm{st}}(x)\propto e^{-U\left(x\right)/D}$.
For a harmonic potential, $U(x) = \mu x^2/2$, a case which is of particular interest, not only theoretically but also experimentally  \cite{Takatori,Dauchot2019}, the stationary distribution \eqref{eq:Pst_2st} simplifies to \cite{Dhar_2019, TC08}
\be
\label{eq:2st_harmonic}
P_\textrm{st}(x)= \frac{2 \mu}{4^{\beta}B(\beta, \beta)v_0} \left [1- \left(\frac{\mu x}{v_0} \right)^2 \right]^{\beta -1} ,
\ee
where $\beta = \gamma/\mu$ and $B(u,v)$ is the beta-function. The distribution is symmetric, $P_{\textrm{st}}(x)=P_{\textrm{st}}(-x)$ and describes a particle that is confined to the region $\left|x\right|\le v_0 / \mu$.
As one varies $\beta$, the shape of the distribution changes from a unimodal distribution centered around $x=0$ at $\beta > 1$, describing a `passive phase', to a bimodal distribution in which the peaks are near the edges $x = \pm v_0 / \mu$ at $\beta < 1$, describing an `active phase' (at $\beta=1$ the distribution is uniform).
In the strongly passive limit $\beta \gg 1$, the distribution \eqref{eq:2st_harmonic} becomes a Gaussian $P_{\textrm{st}}(x)\propto e^{-\gamma\mu x^{2}/v_{0}^{2}}$, corresponding to a Boltzmann distribution with diffusivity $D = v_0^2/2\gamma$.

In fact, for a harmonic potential, the full time-dependent position distribution $P(x,t)$ has recently been obtained exactly \cite{Dhar_2019}. It is given, in terms of its Laplace transform
\be
\tilde{P}\left(x,s\right)=\int_{0}^{\infty}e^{-st}P\left(x,t\right)dt \, ,
\ee
by
\be
\label{Ptildesol}
\tilde{P}\left(x,s\right)=B\left(s\right)z^{\bar{\gamma}+\bar{s}-1}\,_{2}F_{1}\left(1-\bar{\gamma},\bar{\gamma};\bar{\gamma}+\bar{s};z\right) \, ,
\ee
where
\bea
\label{Bsdef}
B\left(s\right)&=&2^{2\left(\bar{\gamma}+\bar{s}\right)-3}\frac{\Gamma\left(\bar{s}/2\right)\Gamma\left[\bar{\gamma}+\left(1+\bar{s}\right)/2\right]}{\sqrt{\pi}\Gamma\left(\bar{\gamma}+\bar{s}\right)} \,,\\
z&=&\frac{1}{2}\left(1-\frac{\mu |x|}{v_{0}}\right) \, ,
\eea
$\bar{s}=s/\mu$, $\bar{\gamma}=\gamma/\mu$, and $\,_{2}F_{1}(a,b;c;d)$ is a standard hypergeometric function.
One can check that in the free case $\mu=0$, the result simplifies to Eq.~\eqref{p1xt.1}.

The principal goal of the present work is to calculate the position distribution of an RTP in higher spatial dimension, focusing mostly on two dimensions (2D), thereby extending the known 1D results. Such extensions are very important from the point of view of relevance to experiments.
In contrast to 1D where there is essentially just one natural definition of an RTP whose speed is constant, in 2D different models of active particles have been introduced and studied, with growing interest over the last few years \cite{ABP2018, Malakar20, Basu20, SBS20, MB20, SBS21, CO21, Frydel22, SBS22, Angelani22, MP22, NG22, CF22}.

In the active Brownian particle (ABP) model, the velocity can be oriented toward any direction in the plane, and the orientation changes continuously in time through angular diffusion.
The position distribution of an ABP in 2D was studied, with and without an external harmonic potential, in Refs. \cite{ABP2018, ABP2019, Malakar20, NG22, CF22}.
There are different RTP models in 2D (with an orientation that changes discontinuously in time), in which the details differ: The set of possible orientations can be finite or infinite, and different possible transition rules of the orientation have been studied.
The steady-state distribution of a RTP whose orientation is chosen randomly at each tumbling event, uniformly from all possible orientations, in the presence of a harmonic trap in 2D and 3D, was obtained very recently in \cite{Frydel22}.
In the present work, we significantly advance the understanding of RTPs in 2D by finding the exact, time-dependent position distribution of an RTP in 2D whose orientation vector stochastically rotates by 90 degrees (clockwise or counter-clockwise), confined by an external harmonic potential and possibly diffusing, as well as some extensions of this model.
This is achieved by employing a change of coordinates, that leads to a decomposition of the problem  into two statistically-independent one-dimensional problems whose solution has been recently found.

Let us briefly describe the structure of the remainder of the paper. 
In section \ref{sec:model}, we give the precise definition of the 2D RTP model that we study.
In section \ref{sec:solution}, we solve the model exactly, and calculate the distribution of the position of the particle and related quantities.
In section \ref{sec:extensions}, we present several generalizations of the model (to more than one RTP, to higher dimensions etc) and briefly describe how to extend our results to cover those cases too.
In section \ref{sec:discussion}, we summarize and discuss our main findings.
Some of the technical details of the calculations are given in the Appendices.

\section{Model}
\label{sec:model}

\begin{figure}[ht]
\includegraphics[width=0.5\linewidth,clip=]{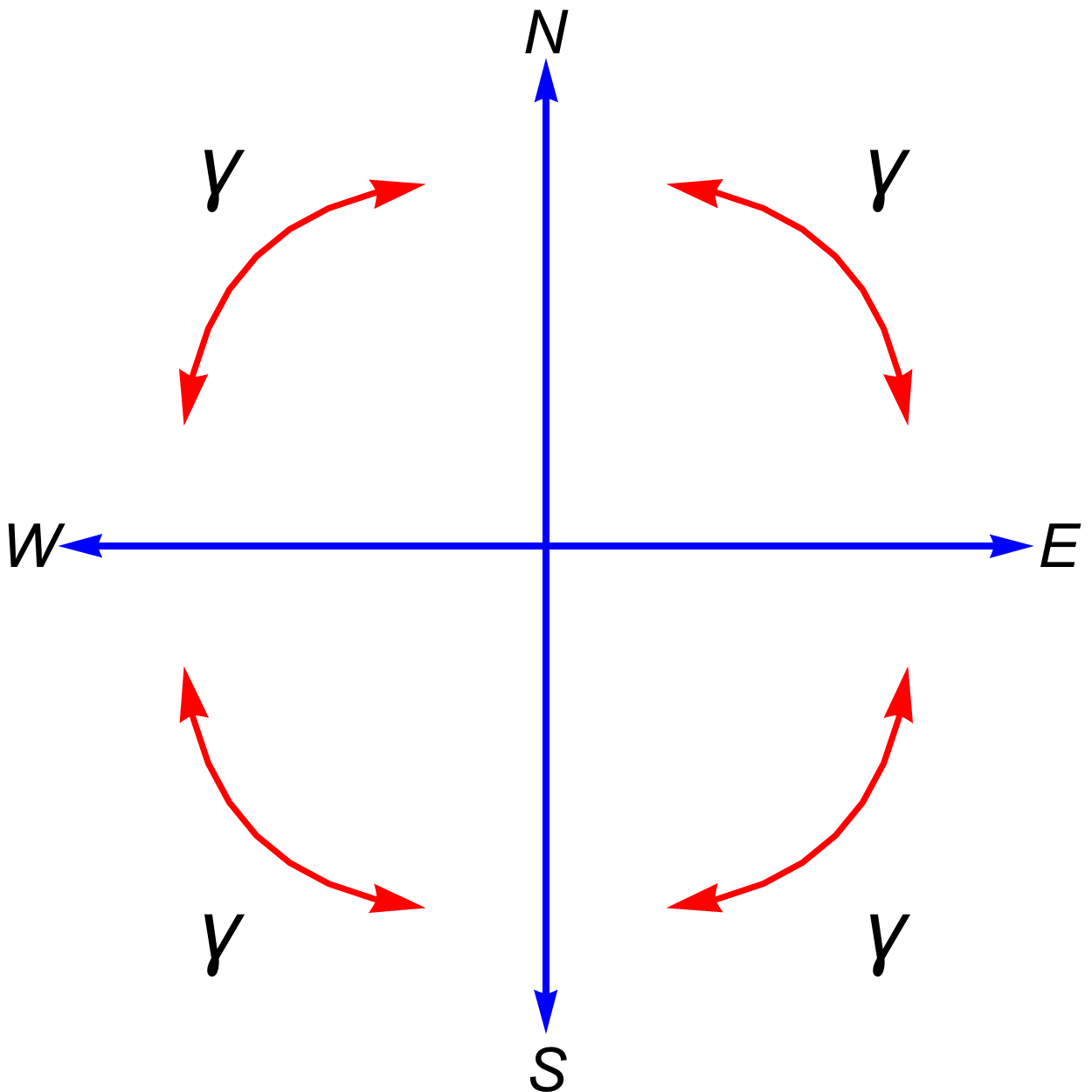}
\caption{A schematic representation of the dynamics of the orientation vector $\vect{\sigma} (t)$ in the 2D RTP model. $\vect{\sigma}$ rotates by 90 degrees to the left or to the right, each at rate $\gamma$. The four possible orientations are aligned with the $x$ and $y$ axes. As explained in the text, the key to our solution of this model is the observation that a 90-degree rotation of $\vect{\sigma}$ corresponds to an inversion of exactly one of its two components $\sigma_{u}$ and $\sigma_{v}$, where $\hat{u}=\left(\hat{x}+\hat{y}\right)/\sqrt{2}$ and $\hat{v}=\left(\hat{x}-\hat{y}\right)/\sqrt{2}$ are coordinates rotated by 45 degrees with respect to the $x,y$ coordinates. In fact, we find that $\sqrt{2}\,\sigma_{u}(t)$ and $\sqrt{2}\,\sigma_{v}(t)$ are statistically-independent telegraphic noises, leading to a complete decoupling of the problem in the $u,v$ coordinates.}
\label{fig:SigmaDynamics}
\end{figure}

The 2D RTP model that we study was originally introduced in \cite{Basu20}. 
It consists of an overdamped particle in the 2D ($xy$) plane, which is is affected by an external harmonic potential  $U\left(x,y\right)=\mu\left(x^{2}+y^{2}\right)/2$, and in addition, has an internal degree of freedom $\vect{\sigma}$ that is a unit vector, describing the particle's orientation.
The dynamics of the particle's position are described by the Langevin equation
\be
\label{rdot}
\dot{\vect{r}}=-\mu \vect{r}\left(t\right)+\sqrt{2}\,v_{0}\vect{\sigma}\left(t\right)\,,
\ee
where $\vect{r}=\left(x,y\right)$ is the position of the particle and $\sqrt{2} \, v_{0}$ would be the particle's speed in the absence of external potential (the factor $\sqrt{2}$ being included for later convenience). 
The dynamics of $\vect{\sigma}$ are stochastic: it rotates (``tumbles'') by 90 degrees, clockwise or counter-clockwise with equal probabilities, each of the rotations occurring at a constant rate $\gamma$. Thus, there are four possible orientations for $\vect{\sigma}$, which we choose to be in the directions 
$\pm\hat{x},\pm\hat{y}$ (where $\hat{x}$ and $\hat{y}$ are unit vectors in the directions of the $x$ and $y$ axes, respectively). These four directions are denoted by $E,W,N,S$ respectively, see Fig.~\ref{fig:SigmaDynamics}.
Thus, the master equation that describes the dynamics of $\vect{\sigma}$ is
\be
\label{MasterEq}
\frac{d}{dt}\left(\begin{array}{c}
p_{E}\\
p_{N}\\
p_{W}\\
p_{S}
\end{array}\right)=\gamma\left(\begin{array}{cccc}
-2 & 1 & 0 & 1\\
1 & -2 & 1 & 0\\
0 & 1 & -2 & 1\\
1 & 0 & 1 & -2
\end{array}\right)\left(\begin{array}{c}
p_{E}\\
p_{N}\\
p_{W}\\
p_{S}
\end{array}\right) \,,
\ee
where $p_i(t)$ denotes the probability that at time $t$, the orientation of the particle is $i=E,N,W,S$.
The particle is initially at the origin, $x\left(t=0\right)=y\left(t=0\right)=0$, with $\vect{\sigma}(t=0)$ uniformly distributed over the 4 possible orientations \cite{Basu20footnote}.

In \cite{Basu20}, the exact steady-state marginal distribution along the $x$ axis was calculated for this model.
However, the relaxation to this steady state, as described by the time dependent distributions, has not been known, and more importantly, neither has the full (two-dimensional) distribution $\mathcal{P}\left(x,y,t\right)$. In particular, the steady-state distribution
$\mathcal{P}_{\text{st}}\left(x,y\right)$
has not been known.
Note that the time-dependent distribution is meaningful and important also for a ``free'' RTP, i.e., in the absence of an external potential, $\mu = 0$ \cite{OK96, Orsingher2000, SBS20, CO21}.
%
In this paper we resolve exactly these outstanding issues:
We find the exact time-dependent distribution of the particle's position $\mathcal{P}\left(x,y,t\right)$ for $\mu \ge 0$, and in particular, we find the steady state distribution $\mathcal{P}_{\text{st}}\left(x,y\right)$ that is reached at long times for $\mu > 0$.
We also solve related problems such as the survival and first-passage properties, and discuss several extensions of the model.

\section{Exact solution}
\label{sec:solution}

The 2D RTP model becomes considerably simpler to analyze if one changes to a coordinate system
\be
\label{uvdef}
u=\frac{x+y}{\sqrt{2}},\quad v=\frac{x-y}{\sqrt{2}},
\ee
 that is rotated by 45 degrees with respect to the $x,y$ coordinates.
In this section, we exactly solve the 2D model using the following three key ingredients:
(i) We show that the $u$ and $v$ components of the noise $\vect{\sigma}$ are statistically-independent telegraphic noises.
(ii) We find that the $u$ and $v$ coordinates of the particle are also statistically-independent processes, and that each of them is mathematically equivalent to the position of a 1D harmonically-confined RTP.
(iii) We recall the known exact results for the 1D case, and employ them to get the solution to the 2D model.
%
%
After obtaining the exact solution, we study some of its properties such as its anisotropy and the relaxation to the steady state. 

\subsection{Decoupling of the noise}

The first key step to solving this model exactly is to observe that $\sqrt{2} \, \sigma_{u}\left(t\right)$ and $\sqrt{2}\, \sigma_{v}\left(t\right)$ (where $\sigma_u$ and $\sigma_v$ are the $u$ and $v$ components of $\vect{\sigma}$) are two statistically-independent telegraphic noises, each of which takes the values $\pm 1$ and stochastically flips its sign with rate $\gamma$. 
%
Let us begin by showing that $\sqrt{2} \, \sigma_{u}\left(t\right)$ is a telegraphic noise. This is quite easy. One simply has
\bea
p_{+}&\equiv&\text{Prob}\left(\sqrt{2}\sigma_{u}=1\right)=p_{E}+p_{N} \, ,\\
p_{-}&\equiv&\text{Prob}\left(\sqrt{2}\sigma_{u}=-1\right)=p_{W}+p_{S} \, .
\eea
By summing the first two components and the last two components of the master equation \eqref{MasterEq}, one then finds that the dynamics of $\left(p_{+},p_{-}\right)$ is governed by the master equation
\bea
\frac{dp_{+}}{dt}&=&-\gamma p_{+}+\gamma p_{-} \, ,\\
\frac{dp_{-}}{dt}&=&-\gamma p_{-}+\gamma p_{+} \, ,
\eea
which coincides exactly with that of a telegraphic noise.
One similarly proves that $\sqrt{2} \, \sigma_{v}\left(t\right)$ is a telegraphic noise.
However, proving that the two processes $\sqrt{2} \, \sigma_{u}\left(t\right)$ and $\sqrt{2}\, \sigma_{v}\left(t\right)$ are statistically independent is a little more tricky.

Let us now prove the statistical independence of the two processes $\sqrt{2} \, \sigma_{u}\left(t\right)$ and $\sqrt{2}\, \sigma_{v}\left(t\right)$, which turns out to be crucial for the solution of the 2D model. Our strategy in the proof is to define a 2D noise $\vect{\Sigma}$ whose $u$ and $v$ components are statistically-independent telegraphic noises, and then, to show that $\vect{\Sigma}$ and $\vect{\sigma}$ are equivalent.
So, let us define a 2D noise
\be
\vect{\Sigma}\left(t\right)=\Sigma_{u}\left(t\right)\hat{u}+\Sigma_{v}\left(t\right)\hat{v} \, ,
\ee
where $\Sigma_{u}\left(t\right)$ and $\Sigma_{v}\left(t\right)$, the $u$ and $v$ components (respectively) of $\vect{\Sigma}\left(t\right)$, are two statistically-independent (decoupled) telegraphic noises, each taking the values $\pm1/\sqrt{2}$ and switching between them at rate $\gamma$. $\Sigma_{u}\left(t=0\right)$ and $\Sigma_{v}\left(t=0\right)$ are each randomly and independently selected from the two possible values that each of them can take. Thus, $\vect{\Sigma}(t=0)$ takes each of the four possible values $\pm \hat{x},  \pm \hat{y}$, each with probability $1/4$ [just like $\vect{\sigma}(t=0)$].

\begin{figure*}[ht]
\includegraphics[width=0.47\linewidth,clip=]{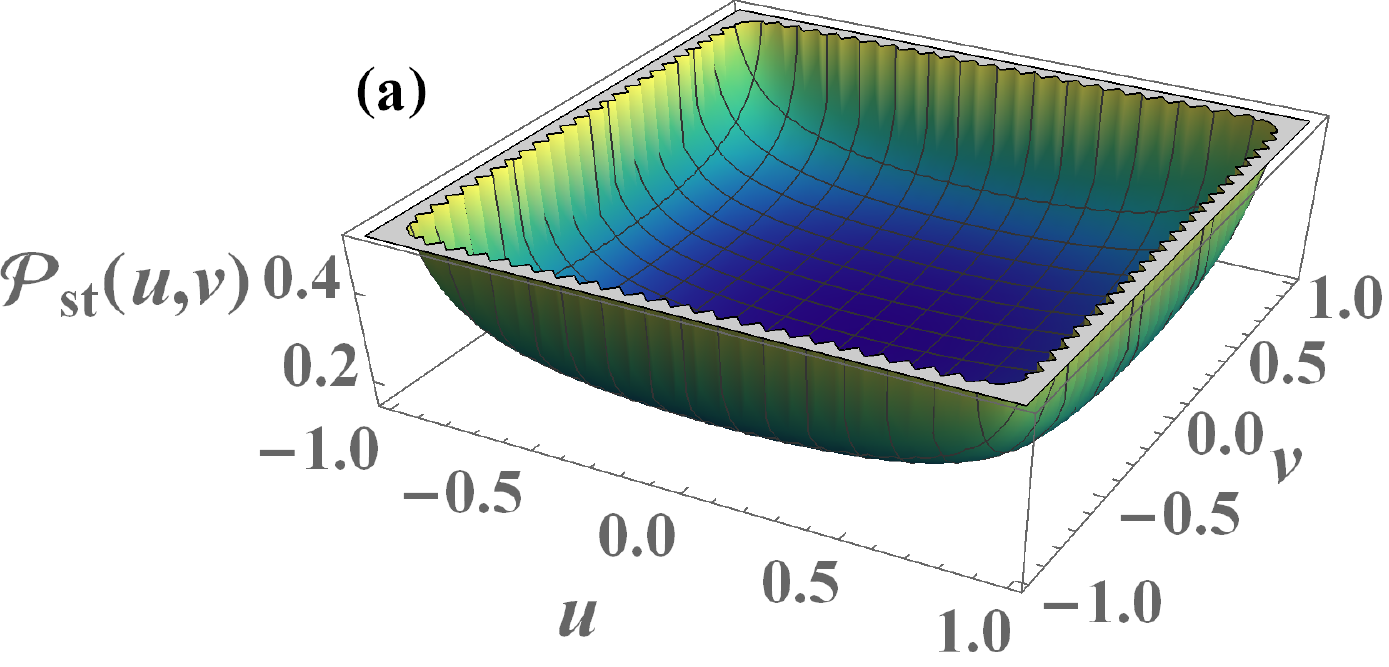}
\hspace{2mm}
\includegraphics[width=0.47\linewidth,clip=]{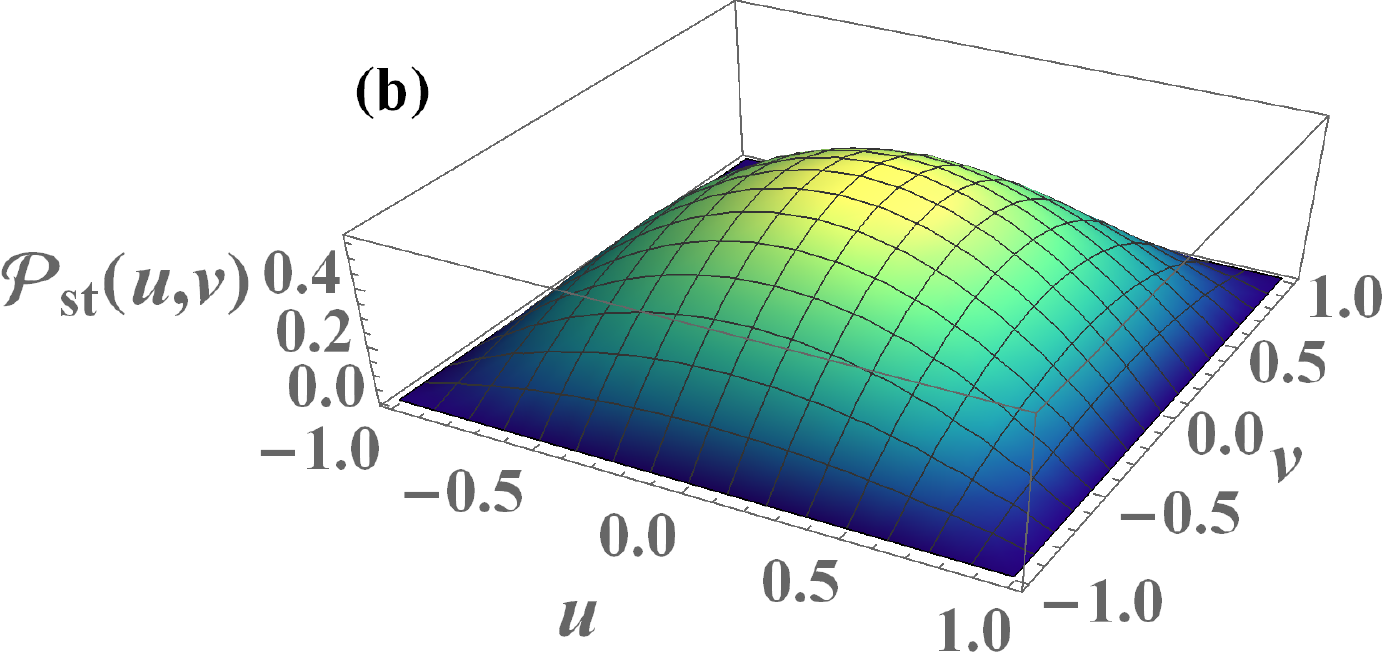}
\caption{The steady-state distribution \eqref{Pstsol} for a 2D RTP confined by a harmonic potential, for $\beta = 2/3$ (a) and $\beta=2$ (b). In this figure, units are chosen such that $\mu = v_0 = 1$. At $\beta < 1$ the particle is most likely to be accumulated away from the center, whereas at $\beta > 1$ it is maximal around the center of the trap. }
\label{fig:Pst}
\end{figure*}

We now show the equivalence between the two processes $\vect{\Sigma}(t)$ and  $\vect{\sigma}(t)$.
Both of them are stationary Markov processes: For $\vect{\Sigma}(t)$, this property is inherited from $\Sigma_{u}\left(t\right)$ and $\Sigma_{v}\left(t\right)$.
Next, we notice that they are both unit vectors, taking one of the four values $\pm\hat{u},\pm\hat{v}$.
In order to show the equivalence between $\vect{\Sigma}(t)$ and  $\vect{\sigma}(t)$, it thus remains to show that the transition rates between these possible values are identical for the two processes, or equivalently, that their dynamics are governed by the same master equation.
Indeed, one finds that the master equation for $\vect{\sigma}$, Eq.~\eqref{MasterEq}, describes the dynamics of $\vect{\Sigma}(t)$ too. A change of sign of $\Sigma_{u}\left(t\right)$ corresponds to one of the transitions $E \leftrightarrow S$ and $W \leftrightarrow N$, while a change of sign of $\Sigma_{v}\left(t\right)$ corresponds to one of the transitions $E \leftrightarrow N$ and $W \leftrightarrow S$, and each of these transitions occurs at rate $\gamma$, leading to the master equation \eqref{MasterEq}.
This completes the proof of the equivalence between $\vect{\Sigma}(t)$ and  $\vect{\sigma}(t)$.
It follows that $\sqrt{2} \, \sigma_{u}\left(t\right)$ and $\sqrt{2}\, \sigma_{v}\left(t\right)$ are two statistically-independent telegraphic noises, each of which takes the values $\pm 1$ and stochastically switches sign at rate $\gamma$.

\subsection{Decoupling of the particle's position}

Now that we have seen that the noise decouples in the $u,v$ coordinates, it is reasonable to expect the analysis of the 2D RTP model to simplify when studied in these coordinates. Indeed,
writing the Langevin dynamics \eqref{rdot} explicitly in the $u,v$ coordinates, we have
\bea
\label{udot}
\dot{u}&=&-\mu u\left(t\right)+\sqrt{2} \, v_{0}\sigma_{u}\left(t\right) \, ,\\
\label{vdot}
\dot{v}&=&-\mu v\left(t\right)+\sqrt{2} \, v_{0}\sigma_{v}\left(t\right) \, .
\eea
One immediately observes that Eqs.~\eqref{udot} and \eqref{vdot} are decoupled, which makes the solution far simpler. Note that no coupling enters through the noise terms, since we have already shown that $\sqrt{2}\,\sigma_{u}\left(t\right)$ and $\sqrt{2}\,\sigma_{v}\left(t\right)$ are statistically-independent telegraphic noises.
Moreover, Eqs.~\eqref{udot} and \eqref{vdot} are mathematically equivalent to the equations that describe two noninteracting, harmonically-confined RTP's in 1D. As a result, $u(t)$ and $v(t)$ are two \emph{statistically-independent} processes, each of which corresponds to a 1D RTP whose free velocity is given by $v_{0}$.

 This decoupling enables us to immediately solve the 2D model exactly, as we now explain. Consider the Green's function 
$\mathcal{P}_{\vect{\sigma}\vect{\sigma}'}\left(\vect{r},t\;|\;\vect{r}\text{’},t\text{’}\right)$ 
that gives the joint distribution of the RTP's position $\vect{r}$ and orientation $\vect{\sigma}$ at time $t$, conditioned on their values at time $t' < t$. We find that the Green's function decomposes (in the $u,v$ coordinates) as
\be
\label{GreenDecomposition}
\mathcal{P}_{\vect{\sigma}\vect{\sigma}'}\left(\vect{r},t\;|\;\vect{r}\text{’},t\text{’}\right)=P_{\sigma_{u}\sigma_{u}'}\left(u,t\;|\;u\text{’},t\text{’}\right)P_{\sigma_{v}\sigma_{v}'}\left(v,t\;|\;v\text{’},t\text{’}\right) \,,
\ee
where
$P_{\sigma\sigma'}\left(x,t\;|\;x\text{’},t\text{’}\right)$
is the Green's function for an RTP in 1D. 
Eq.~\eqref{GreenDecomposition} follows immediately from the arguments given above. However, as an alternative approach, we also recover Eq.~\eqref{GreenDecomposition} by analyzing the joint Fokker-Planck equation for the position and orientation of the particle in Appendix \ref{app:Green}, providing a useful check of this result.

Similarly, the time-dependent position distribution decomposes as
\be
\label{Pxyt}
\mathcal{P}\left(\vect{r},t\right)=P\left(u,t\right)P\left(v,t\right) \, ,
\ee
where $P(\dots,t)$ is the position distribution of a 1D harmonically-confined RTP, and is given above. 
In particular, the steady-state distribution is given by
\be
\mathcal{P}_{\text{st}}\left(u,v\right)=P_{\text{st}}\left(u\right)P_{\text{st}}\left(v\right)
\ee
where $P_{\text{st}}\left(\dots\right)$  is given by Eq.~\eqref{eq:2st_harmonic}. 
Explicitly, the steady-state distribution is
\bea
\label{Pstsol}
&&\mathcal{P}_{\text{st}}\left(u,v\right)=\frac{4\mu^{2}}{2^{4\beta}\left[B\left(\beta,\beta\right)v_{0}\right]^{2}} \nn\\
&&\qquad \times \left[\left(1-\left(\frac{\mu u}{v_{0}}\right)^{2}\right)\left(1-\left(\frac{\mu v}{v_{0}}\right)^{2}\right)\right]^{\beta-1} \, ,
\eea
where we recall that $\beta=\gamma/\mu$,
see Fig.~\ref{fig:Pst}.
The support of the distribution $\mathcal{P}_{\text{st}}\left(u,v\right)$ is the square $\left|u\right|,\left|v\right|<v_{0}/\mu$.
In the active phase $\beta < 1$, the position of the particle accumulates near the edges of the support, the distribution becoming localized around the corners of the square $\left|u\right|,\left|v\right|<v_{0}/\mu$ in the limit $\beta \ll 1$. 
In the passive phase $\beta > 1$ the distribution is maximal near the center of the trap, and in the diffusive limit $\beta \gg 1$, typical fluctuations are described by an isotropic Gaussian distribution around the origin. This corresponds to the passive limit in which the noise can be approximated as white.
For a free particle ($\mu=0$), the distribution never reaches a steady state. However, the time-dependent distribution simplifies, since Eq.~\eqref{Pxyt} reduces to
\be
\label{Pfreesol}
\mathcal{P}_{\text{free}}\left(u,v,t\right)=P_{\text{free}}\left(u,t\right)P_{\text{free}}\left(v,t\right) \, ,
\ee
where $P_{\text{free}}$ is given by \eqref{p1xt.1}.
The solution \eqref{Pfreesol} for the free case has been known for some time \cite{OK96, Orsingher2000, CO21}.

As a useful check of these results, one can calculate the marginal distribution of $x=\left(u+v\right)/\sqrt{2}$ and compare it to the previously-known results \cite{Basu20, SBS20}. In Appendix \ref{app:marginal} we perform this check explicitly for the stationary distribution \eqref{Pstsol} for $\beta\in\left\{ 1,2\right\}$, and for the time-dependent distribution \eqref{Pfreesol} for a free particle, and find perfect agreement.

\subsection{Anisotropy of the distribution}

\begin{figure}[ht]
\includegraphics[width=0.98\linewidth,clip=]{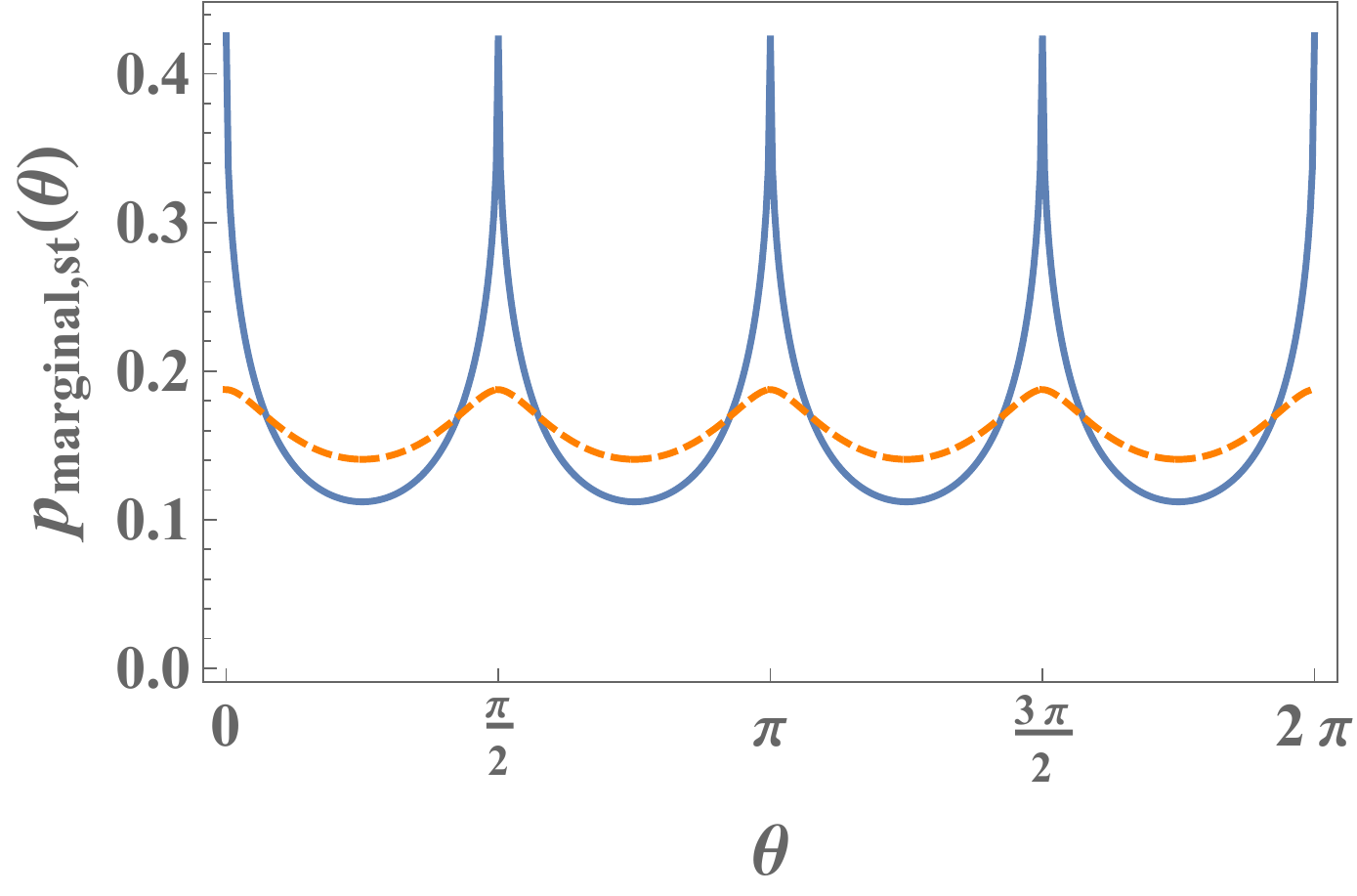}
\caption{The marginal steady-state distribution \eqref{ptheta} of the polar angle $\theta$ of the RTP's position for $\beta = 2/3$ (solid line) and $\beta = 2$ (dashed line). 
}
\label{fig:theta}
\end{figure}

In order to quantify the anisotropy of the distribution, one can consider the marginal distribution of the polar coordinate $\theta$ of the particle's position in the $xy$ plane. In the steady state, this distribution is given by
\bea
\label{ptheta}
&& p_{\text{marginal,st}}\left(\theta\right)\nn\\
&&=\int_{0}^{\infty}\mathcal{P}_{\text{st}}\left(u=r\cos\left(\theta+\frac{\pi}{4}\right),v=-r\sin\left(\theta+\frac{\pi}{4}\right)\right)rdr \nn\\
&& = \int_{0}^{M\left(\phi\right)}\frac{4\left[\left(1-\left(r\cos\phi\right)^{2}\right)\left(1-\left(r\sin\phi\right)^{2}\right)\right]^{\beta-1}}{2^{4\beta}\left[B\left(\beta,\beta\right)\right]^{2}}rdr \, , \nn\\
\eea
where
\be
M\left(\phi\right)\equiv\min\left\{ \left|\frac{1}{\cos\phi}\right|,\left|\frac{1}{\sin\phi}\right|\right\}  \, ,
\ee
and $\phi = \theta + \pi /4$.
It is independent of $\mu$ and $v_0$ (as one could expect from dimensional analysis).
For certain values of $\beta$, the integral can be solved, yielding for instance
\bea
&& \left.p_{\text{marginal,st}}\left(\theta\right)\right|_{\beta=1}=\frac{M^{2}\left(\phi\right)}{8},\\
&& \left.p_{\text{marginal,st}}\left(\theta\right)\right|_{\beta=2} \nn\\
&&=\frac{3M^{2}\left(\phi\right)}{256}\left[M^{4}\left(\phi\right)\left(1-\cos\left(4\phi\right)\right)-12M^{2}\left(\phi\right)+24\right] . \nn\\
\eea
$p_{\text{marginal,st}}\left(\theta\right)$ is plotted in Fig.~\ref{fig:theta} for $\beta = 2/3$ and $\beta=2$.
$p_{\text{marginal,st}}\left(\theta\right)$ is maximal (as a function of $\theta$) in the directions of the possible orientations of the noise, 
$\theta\in\left\{ 0, \pi/2,\pi, 3\pi/2\right\} $,
 and minimal in the directions of the $u$ and $v$ axes.
In the active limit $\beta \ll 1$, the anisotropy becomes very pronounced, because the distribution is localized around the corners of the square $\left|x\right|,\left|y\right|<v_{0}/\mu$.
In the opposite (diffusive) limit, $\beta \gg 1$, the anisotropy becomes very weak (as explained above), i.e., $p_{\text{marginal,st}}\left(\theta\right)$ is nearly uniform on the interval $0<\theta<2\pi$.
These limiting behaviors are not shown in the figure.

\subsection{Relaxation to the steady state for a generic initial orientation distribution}

One can easily extend the discussion to the slightly more general problem of a 2D harmonically-confined RTP initially at the origin $x\left(t=0\right)=y\left(t=0\right)$
whose initial orientation is given by some distribution
\be
\label{OrientationDist}
\left(p_{E}^{\left(0\right)},p_{N}^{\left(0\right)},p_{W}^{\left(0\right)},p_{S}^{\left(0\right)}\right)
\ee
where $p_{i}^{\left(0\right)}=p_{i}\left(t=0\right)$ is the probability that the initial orientation is in the direction $i$.
For a deterministic initial orientation, the position distribution $\mathcal{P}\left(\vect{r},t\right)$ will decompose in the $u,v$ coordinates (this can be seen, for instance, by summing Eq.~\eqref{GreenDecomposition} with $\vect{r}' = t' = 0$ over the four possible values of $\vect{\sigma}$).
This result can be easily extended to some general initial orientation distribution \eqref{OrientationDist} by using the superposition principle, and the result will simply be
\bea
\label{PuvtOrientationDist}
&& \mathcal{P}\left(u,v,t\right)=\sum_{\sigma_{1}=\pm}\sum_{\sigma_{2}=\pm}p_{\sigma_{u},\sigma_{v}}^{\left(0\right)} \nn\\
&& \quad \times  \, P\left(u,t\,|\,\sigma_{u}\left(t=0\right)=\sigma_{1}\right)P\left(v,t\,|\,\sigma_{v}\left(t=0\right)=\sigma_{2}\right) \,, \nn\\
\eea
where, in the expression $p_{\sigma_{u},\sigma_{v}}^{\left(0\right)}$, we identify the four possible orientations of the noise with their corresponding $u$ and $v$ components of the noise, i.e.,
\be
E\equiv\left(+,+\right),\;\;\; N\equiv\left(+,-\right),\;\;\; W\equiv\left(-,-\right),\;\;\; S\equiv\left(-,+\right) ,
\ee
and where
\be
P\left(u,t\,|\,\sigma\left(t=0\right)=\sigma_{0}\right)
\ee
is the position distribution of a 1D harmonically-confined RTP whose initial orientation is $\sigma_0$.
Eq.~\eqref{PuvtOrientationDist} describes the relaxation of the position distribution to the steady state \eqref{Pstsol}, which is reached in the long-time limit $t \to \infty$ for any initial condition.

\subsection{First-passage and survival properties}

The statistical independence of $u(t)$ and $v(t)$ has additional important consequences, beyond the decomposition \eqref{Pxyt} of their joint distribution. 
One such consequence is that survival and exit probabilities for a 2D RTP are related to the corresponding ones in 1D, for certain geometries. 
In 1D, such problems have been studied quite extensively \cite{ang14,ang15,MalakarEtAl18,EM_2018,Maes2018, led19,Singh2019,led20,ban20,mor20, Wexler20, Dhar_2019}.
For instance, the first-passage time $t_{q}$ of the 2D RTP $\vect{r}(t)$ out of the quadrant $\left\{ u>0,v>0\right\} $ is defined as the first time at which the particle exits the quadrant.
Clearly, $t_{q} = \min\left\{ t_{u},t_{v}\right\} $ where $t_u$ and $t_v$ are the first-passage times of $u(t)$ and $v(t)$ out of the half lines $u>0$ and $v>0$ respectively. Now, since $u(t)$ and $v(t)$ are statistically independent, so are $t_u$ and $t_v$.
As a result, the cumulative distribution function $\text{Prob}\left(t_{q}<\tau\right)$ of $t_q$, that describes the probability that the particle `survives' (i.e., remains) inside the quadrant up to time $\tau$ is given by the product of the probabilities that $u(t)$ and $v(t)$ remain positive up to time $\tau$, i.e, 
\bea
\label{survival2D}
\text{Prob}\left(t_{q}<\tau\right)&=&\text{Prob}\left(t_{u}<\tau\right)\text{Prob}\left(t_{v}<\tau\right)=\nn\\&=&\left[\text{Prob}\left(t_{u}<\tau\right)\right]^{2} \, ,
\eea
where in the second equality we used the fact that $t_u$ and $t_v$ are identically distributed.
The distribution of $t_u$ is exactly known for the free case $\mu=0$, see Refs.~\cite{MalakarEtAl18, led19}, and using this result together with Eq.~\eqref{survival2D}, one obtains the distribution of $t_q$ for the free case.

\section{Extensions}
\label{sec:extensions}

In this section we briefly outline some extensions of these results in several directions.

\subsection{Different propulsion velocities}

Let us consider a version of the model in which the $x$ and $y$ directions are not on equal footing. We will assume that the propulsion velocities in the $x$ and $y$ directions are different, by allowing the possible values of the noise term to be $\pm\hat{x},\pm\alpha y$ for some general $\alpha > 0$. 

It turns out that the same method that we used above can be straightforwardly extended to this case, with only minor technical complications, as follows. The noise term still decouples as
\be
\label{sigmauvAnisotropic}
\vect{\sigma}\left(t\right)=\sigma_{u}\left(t\right)\frac{\hat{x}+\alpha\hat{y}}{\sqrt{2}}+\sigma_{v}\left(t\right)\frac{\hat{x}-\alpha\hat{y}}{\sqrt{2}}
\ee
where $\sqrt{2} \sigma_u(t)$ and $\sqrt{2} \sigma_v(t)$ are two statistically-independent telegraphic noises.
Similarly, one expresses the position of the RTP as
\be
\label{ruvAnisotropic}
\vect{r}\left(t\right)=u\left(t\right)\frac{\hat{x}+\alpha\hat{y}}{\sqrt{2}}+v\left(t\right)\frac{\hat{x}-\alpha\hat{y}}{\sqrt{2}}.
\ee
Now, by using Eqs.~\eqref{rdot}, \eqref{sigmauvAnisotropic} and \eqref{ruvAnisotropic} one finds that the dynamics of $u$ and $v$ are given by Eqs.~\eqref{udot} and \eqref{vdot} respectively.
As a result, $u(t)$ and $v(t)$ are statistically-independent, and each of them mathematically corresponds to the position of a 1D harmonically-confined RTP. Therefore, the position distribution $\mathcal{P}\left(\vect{r},t\right)$ is still straightforward to find, with only a small technical complication arising from the fact that the basis 
$\left\{ \frac{\hat{x}+\alpha\hat{y}}{\sqrt{2}},\frac{\hat{x}-\alpha\hat{y}}{\sqrt{2}}\right\} $ 
is not orthonormal for general $\alpha$.
Using the transformation formula
\be
\label{uvalpha}
u=\frac{x+y/\alpha}{\sqrt{2}},\quad v=\frac{x-y/\alpha}{\sqrt{2}},
\ee
we find that Eq.~\eqref{Pxyt} gives way to
\be
\mathcal{P}\left(\vect{r},t\right)=\frac{1}{2}\left(1+\frac{1}{\alpha^{2}}\right)P\left(u,t\right)P\left(v,t\right)\,,
\ee
where the factor $\frac{1}{2}\left(1+\frac{1}{\alpha^{2}}\right)$ is the Jacobian of the transformation \eqref{uvalpha}.

\subsection{Two interacting RTP's}

It is fairly straightforward to extend our results to two 2D RTPs with a harmonic interaction (and possibly confined by an external harmonic potential), as long as their possible orientation vectors are the same.
The Langevin equations describing the time evolution of the positions $\vect{r}_{A}$ and $\vect{r}_{B}$ of the two particles are
\bea
\antiquad \dot{\vect{r}}_{A}&=&-\mu\vect{r}_{A}\left(t\right)-\lambda\left(\vect{r}_{A}\left(t\right)-\vect{r}_{B}\left(t\right)\right)+\sqrt{2}\,v_{0}\vect{\sigma}_{A}\left(t\right), \\
\antiquad \dot{\vect{r}}_{B}&=&-\mu\vect{r}_{B}\left(t\right)-\lambda\left(\vect{r}_{B}\left(t\right)-\vect{r}_{A}\left(t\right)\right)+\sqrt{2}\,v_{0}\vect{\sigma}_{B}\left(t\right),
\eea
where $\lambda$ is the strength of the harmonic interaction between the particles, and $\vect{\sigma}_{A}\left(t\right)$ and $\vect{\sigma}_{B}\left(t\right)$ are two statistically-independent noises, each of which is defined as in the single-particle model \eqref{rdot}.
By rewriting these equations in the $u,v$ coordinates,
one simply finds that the problem decouples into two independent problems in the directions $u$ and $v$, each of which consists of two 1D RTPs with a harmonic interaction, so the joint distribution of the positions of the two RTPs is given by
\be
\mathcal{P}\left(u_{A},v_{A},u_{B},v_{B},t\right)=P\left(u_{A},u_{B},t\right)P\left(v_{A},v_{B},t\right)\,,
\ee 
where $u_A(t)$ and $v_A(t)$ are the $u$ and $v$ coordinates of the first particle (and similarly for the second particle).
The corresponding 1D problem was studied in Ref.~\cite{LMS21}, and the steady state was obtained exactly $P\left(x_{1},x_{2},t\to\infty\right)$ for a general attractive interaction in the case $\mu=0$ (however, our extension to 2D only works if the interaction is harmonic).

\subsection{General damping strength:}
One can extend this model to an RTP that is not (strongly) overdamped, by taking into account an additional inertial term $m \ddot{\vect{r}}$ in Eq.~\eqref{rdot}.
The decomposition in the $u$ and $v$ coordinates will still work, i.e., Eq.~\eqref{Pxyt} will still hold.
The 1D distribution $P(x,t)$ is not exactly known in the presence of an inertial term. It is, however, known in the limit of zero damping, for the free case $\mu = 0$ \cite{Dean21}.
Moreover, for $\mu > 0$ the corresponding steady-state distribution $P_{\text{st}}\left(x\right)$ is approximately known in various limits, such as the rapidly-tumbling limit $\gamma \to \infty$, for the RTP and similar models of active particles \cite{BenIsaac15, Wexler20, Woillez20, SmithFarago22}.

\subsection{Diffusion:}
One can further take into account diffusion. In 1D, for instance, one can consider
\be
\label{RTPDiff1D}
\dot{x}=-\mu x+v_{0}\sigma\left(t\right)+\xi\left(t\right) \, ,
\ee
which is Eq.~\eqref{eq:RTP_2st} for the harmonic force $f(x) = -\mu x$, with an additional white (Gaussian) noise term $\xi(t)$, with zero mean $\left\langle \xi\left(t\right)\right\rangle =0$ and correlation function $\left\langle \xi\left(t\right)\xi\left(t'\right)\right\rangle =2D\delta\left(t-t'\right)$ (here $D$ is the diffusion coefficient, and angular brackets denote ensemble averaging).
Since Eq.~\eqref{RTPDiff1D} is linear, $x(t)$ can be written as the sum of two independent stochastic processes,
\be
x\left(t\right)=x_{1}\left(t\right)+x_{2}\left(t\right) \, ,
\ee
which each follows the original dynamics but with just a single noise term, i.e., 
\bea
\dot{x}_{1}&=&-\mu x_{1}+v_{0}\sigma\left(t\right)\,,\\
\dot{x}_{2}&=&-\mu x_{2}+\xi\left(t\right)\,.
\eea
As a result, the distribution $P(x,t)$ is simply given by the convolution\
\be
\label{convolution}
P\left(x,t\right)=\int_{-\infty}^{\infty}P_{1}\left(x_{1},t\right)P_{2}\left(x-x_{1},t\right)dx_{1}\,,
\ee
where $P_{1}\left(x_{1},t\right)$ and $P_{2}\left(x_{2},t\right)$ are the distributions that correspond to the processes $x_1(t)$ and $x_2(t)$, respectively, and are each exactly known ($x_1(t)$ being the confined 1D RTP studied in \cite{Dhar_2019} as described above, and $x_2(t)$ being an Ornstein-Uhlenbeck process).
This decomposition can be generalized, in 1D, to the sum of any number of noise terms of any type \cite{Tucci22, LMS21}.

Returning to 2D, we could consider \eqref{rdot} with an additional white-noise term,
\be
\label{RTPDiff2D}
\dot{\vect{r}}=-\mu\vect{r}\left(t\right)+\sqrt{2}\,v_{0}\vect{\sigma}\left(t\right)+\vect{\xi}\left(t\right)\,,
\ee
where $\left\langle \vect{\xi}\left(t\right)\right\rangle =0$ and 
$\left\langle \vect{\xi}_{i}\left(t\right)\vect{\xi}_{j}\left(t'\right)\right\rangle =2D\delta_{ij}\delta\left(t-t'\right)$ (for $i,j\in\left\{ x,y\right\} $.
As in the case $D=0$, one finds that the dynamics decouple in the coordinates $u$ and $v$, and in each of these two coordinates one has to consider the 1D dynamics \eqref{RTPDiff1D}. Thus, the decomposition \eqref{Pxyt} is still valid, but with $P(x,t)$ now given by Eq.~\eqref{convolution}.

For this 2D model with diffusion, Eq.~\eqref{RTPDiff2D}, we can calculate the (internal) entropy production rate (EPR), as was recently done in 1D in \cite{GP21}. The EPR is defined as the Kullback–Leibler distance between
forward and backward paths (a precise definition can be found in \cite{GP21}).
They found that the EPR for a 1D diffusing RTP [described by Eq.~\eqref{RTPDiff1D}] is given by
\be
\dot{S}_{\text{1D}}=\frac{2\gamma v_{0}^{2}}{D\left(\mu+2\gamma\right)} \, ,
\ee
assuming that the observer knows the orientation $\sigma(t)$.
Since the 2D model \eqref{RTPDiff2D} decouples in the $u,v$ coordinates, and since the entropy is additive, one simply finds that the EPR in the 2D model is given by $\dot{S}_{\text{2D}}=2\dot{S}_{\text{1D}}$.

\subsection{Stochastic resetting:}
Let us consider the case of a free particle, $\mu=0$, but with stochastic resetting of the position of the particle to the origin at rate $\tilde{r}$ (which is not to be confused with the RTP's radial distance $r$ from the origin). For simplicity, let us assume that when resetting occurs, the orientation is randomly chosen from its 4 possible values with equal probabilities. At long times, the position distribution approaches a nonequilibrium steady state, that we now find exactly. We thus extend the 1D result of \cite{EM_2018}, that is given by
\be
\label{Pfreer1D}
P_{\text{free},\tilde{r}}\left(x\right)=\frac{\lambda\left(\tilde{r}\right)}{2}e^{-\lambda\left(\tilde{r}\right)\left|x\right|} \, \,
\ee
where
\be
\lambda\left(\tilde{r}\right)=\frac{\sqrt{\tilde{r}\left(\tilde{r}+2\gamma\right)}}{v_{0}}\,.
\ee
Before performing the calculation, we notice that $u(t)$ and $v(t)$ are 1D RTP's with stochastic resetting. However, the resetting leads to a statistical dependence between $u$ and $v$. Therefore, their joint distribution is nontrivial, since it is not given by the product of the 1D distributions.

Denoting by $\mathcal{P}_{\text{free},\tilde{r}}\left(\vect{r},t\right)$ the time-dependent position of the particle initially at the origin with a random orientation and with stochastic resetting, we find that \cite{EMS20}
\bea
\label{Prdef}
&&\antiquad \mathcal{P}_{\text{free},\tilde{r}}\left(\vect{r},t\right)= \nn\\
&& e^{-\tilde{r}t}\mathcal{P}_{\text{free},0}\left(\vect{r},t\right)+\tilde{r}\int_{0}^{t}e^{-\tilde{r}\tau}\mathcal{P}_{\text{free},0}\left(\vect{r},\tau\right)d\tau \, ,
\eea
where $\mathcal{P}_{\text{free},0}=\mathcal{P}_{\text{free}}$ is the distribution in the absence of resetting.
Eq.~\eqref{Prdef} follows from a renewal approach, the first term on the right-hand side corresponding to the case in which no resetting events occurs on the time interval $[0,t]$, and in the second term corresponding to the case in which at least one such event occurs, and the integral is over the time $t-\tau$ of the last resetting event before time $t$.
The steady state is obtained by taking the limit $t\to\infty$ in \eqref{Prdef}, and it gives
\be
\label{PstfreeAndLaplace}
\mathcal{P}_{\text{free},\tilde{r}}^{\text{st}}\left(\vect{r}\right)=\tilde{r}\tilde{\mathcal{P}}_{\text{free}}\left(\vect{r},\tilde{r}\right)
\ee
where 
\be
\tilde{\mathcal{P}}_{\text{free}}\left(\vect{r},\tilde{r}\right)=\int_{0}^{\infty}e^{-\tilde{r}t}\mathcal{P}_{\text{free}}\left(\vect{r},t\right)dt
\ee
is the Laplace transform of the time dependent distribution $\mathcal{P}_{\text{free}}\left(\vect{r},t\right)$.

\begin{figure*}[ht]
\includegraphics[width=0.32\linewidth,clip=]{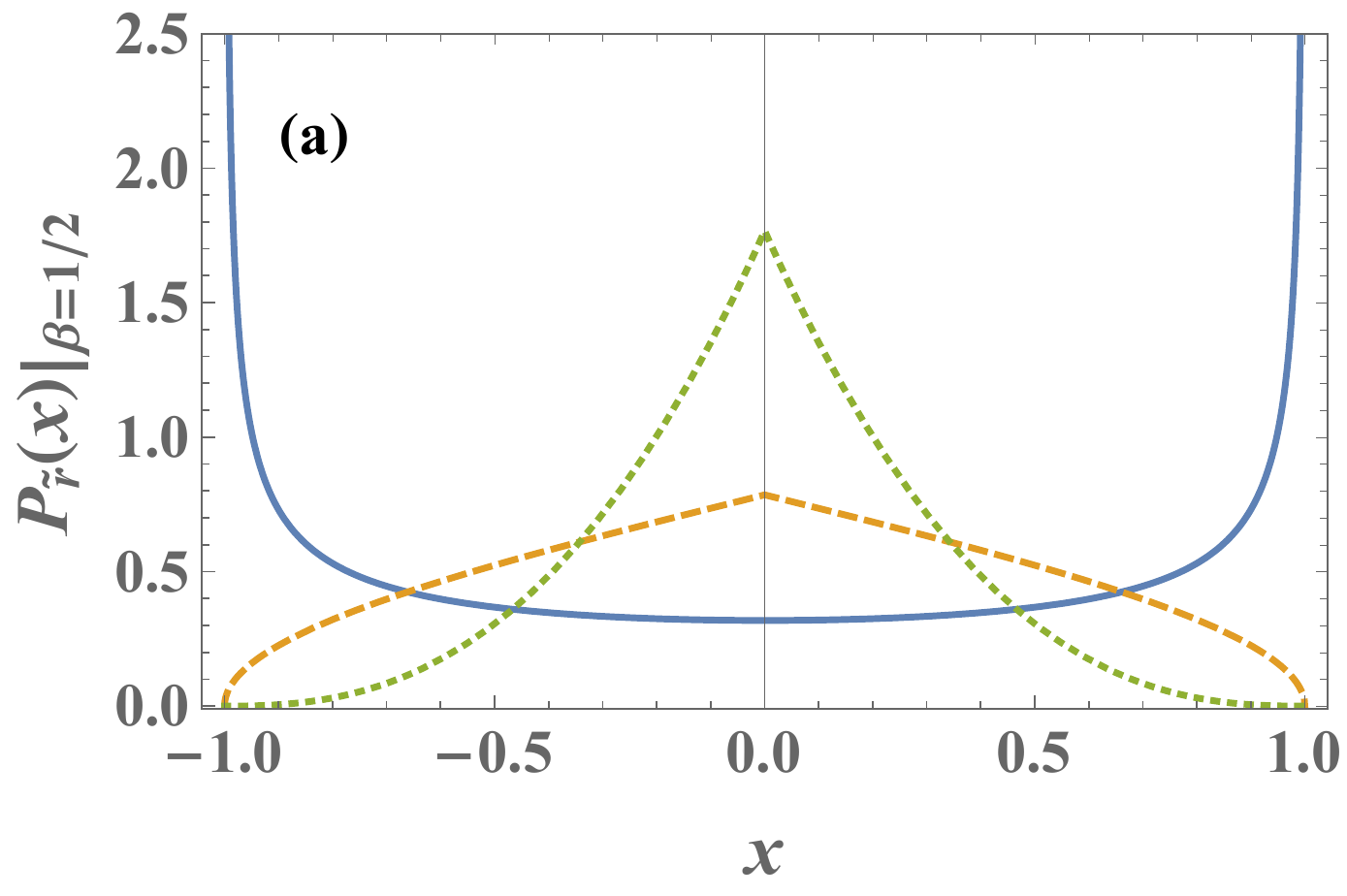}
\hspace{1mm}
\includegraphics[width=0.32\linewidth,clip=]{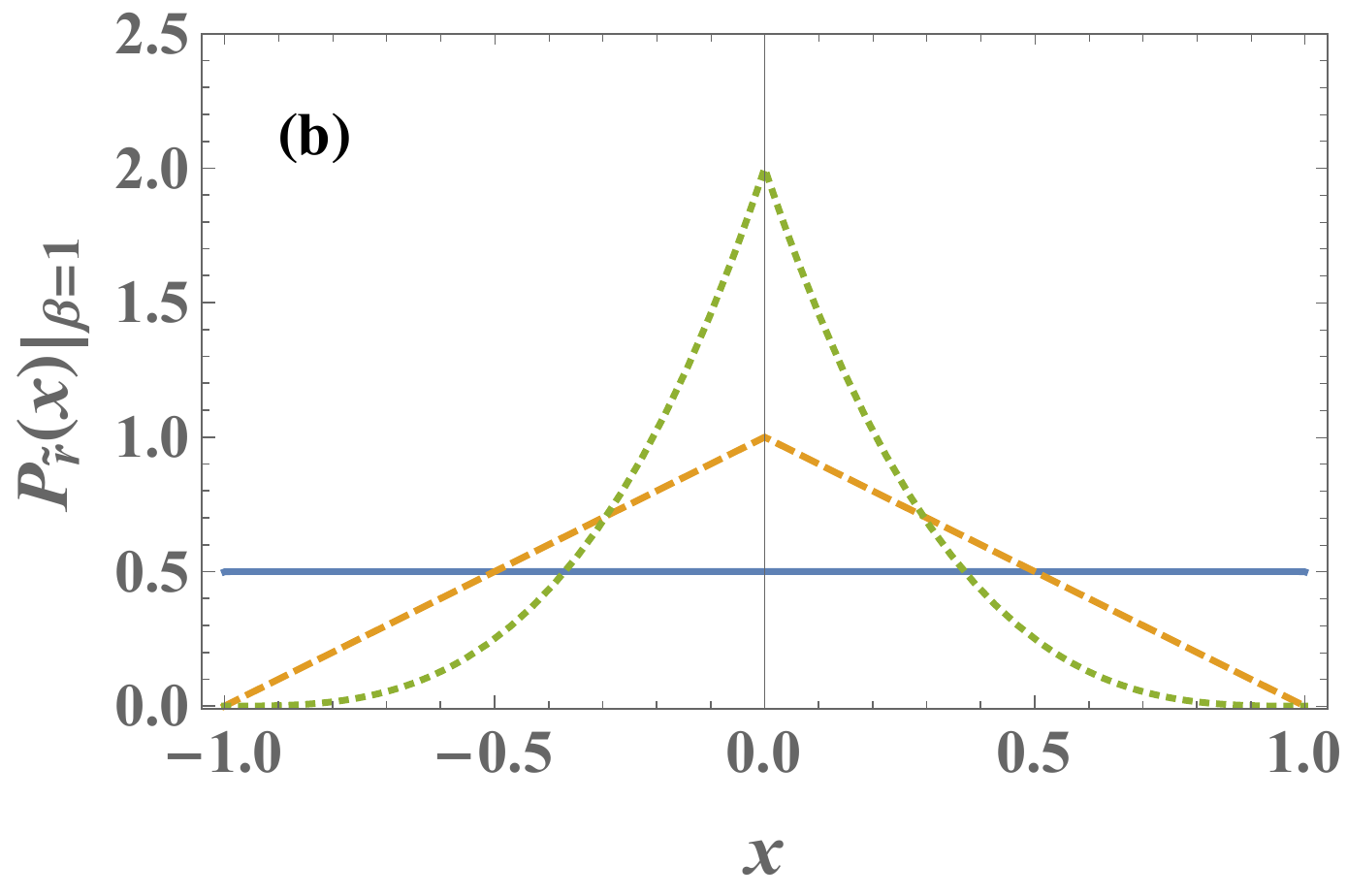}
\hspace{1mm}
\includegraphics[width=0.32\linewidth,clip=]{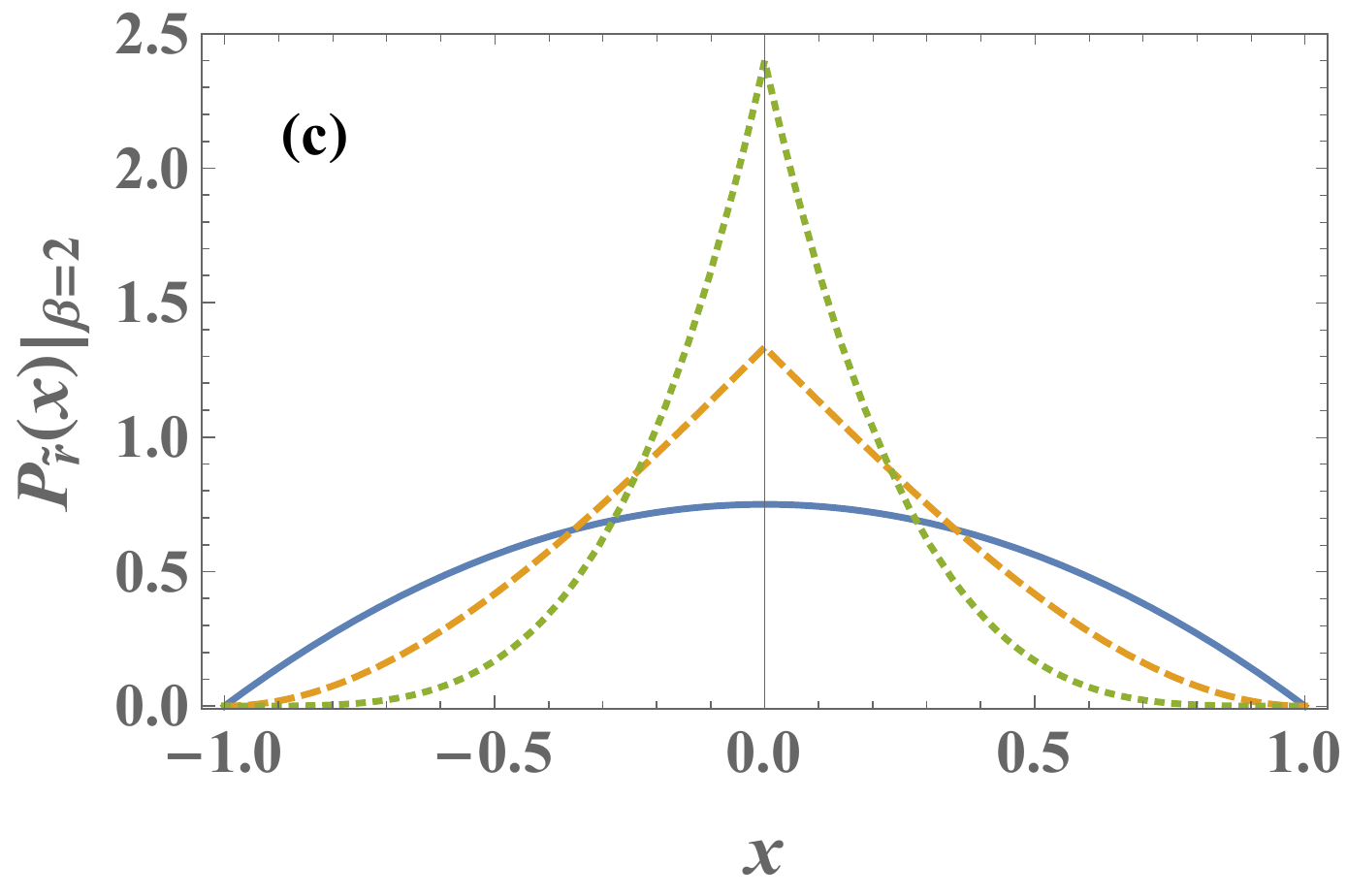}
\caption{The steady-state distribution $P_{\tilde{r}}(x)$ for a 1D RTP confined by a harmonic potential, for $\beta = 1/2$ (a),  $\beta=1$ (b) and $\beta = 2$ (c), with resetting rates $\tilde{r}=0$ (solid lines), $\tilde{r}=1$ (dashed lines) and $\tilde{r}=3$ (dotted lines), see Eq.~\eqref{steadyStateResetting1Dexplicit}.
In this figure, units are chosen such that $\mu = v_0 = 1$. The resetting causes the particle to be localized closer to the center of the trap.}
\label{fig:Prx}
\end{figure*}

It turns out to be convenient to take a Fourier transform in space, i.e., to calculate
\bea
\mathcal{Q}_{\text{free},\tilde{r}}^{\text{st}}\left(\vect{k}\right)&=&\int e^{i\vect{k}\cdot\vect{r}}\mathcal{P}_{\text{free},\tilde{r}}^{\text{st}}\left(\vect{r}\right)d\vect{r}=\nn\\
&=&\int_{0}^{\infty}\tilde{r}e^{-\tilde{r}t}dt\int e^{i\vect{k}\cdot\vect{r}}\mathcal{P}_{\text{free}}\left(\vect{r},t\right)d\vect{r}
\eea
Working in the $u,v$ coordinates, we find, using the decomposition \eqref{Pfreesol} of $\mathcal{P}_{\text{free}}^{\text{st}}\left(\vect{r},t\right)$, that
\be
\label{Qkukvdef}
\mathcal{Q}_{\text{free},\tilde{r}}^{\text{st}}\left(k_{u},k_{v}\right)=\int_{0}^{\infty}\tilde{r}e^{-\tilde{r}t}Q_{\text{free}}\left(k_{u},t\right)Q_{\text{free}}\left(k_{v},t\right)dt\,,
\ee
where 
\be
Q_{\text{free}}\left(k,t\right) = \int_{-\infty}^{\infty}e^{ikx}P_{\text{free}}\left(x,t\right)dx
\ee
is the Fourier transform of the distribution \eqref{p1xt.1} of the position of a free 1D RTP, and is exactly known \cite{MalakarEtAl18} (in the rest of this subsection, we choose units in which $\gamma = v_0 = 1$)
\bea
\label{Pfreek}
&&\!\!\!\!\!\! Q_{\text{free}}\left(k,t\right)=e^{-t} \nn\\
&&\!\!\!\!\!\! \times \left[\cosh\left(t\sqrt{1-k^{2}}\right)+\frac{1}{\sqrt{1-k^{2}}}\sinh\left(t\sqrt{1-k^{2}}\right)\right].
\eea
Plugging \eqref{Pfreek} into \eqref{Qkukvdef}, the integral over $t$ can be performed because the integrand can be written as a sum of exponentials. The result is:
\bea
\label{Qfreersol}
&&\antiquad \mathcal{Q}_{\text{free},\tilde{r}}^{\text{st}}\left(k_{u},k_{v}\right) =\nn\\
&& \frac{\tilde{r}\left(\tilde{r}+2\right)\left[k_{u}^{2}+k_{v}^{2}+\left(\tilde{r}+2\right)\left(\tilde{r}+4\right)\right]}{\left(k_{u}^{2}-k_{v}^{2}\right)^{2}+\left[2\left(k_{u}^{2}+k_{v}^{2}\right)+\tilde{r}\left(\tilde{r}+4\right)\right]\left(\tilde{r}+2\right)^{2}} \, .
\eea
A useful check for the result \eqref{Qfreersol} is that, when plugging in $k_v=0$, we obtain
\be
\mathcal{Q}_{\text{free},\tilde{r}}^{\text{st}}\left(k_{u},k_{v}=0\right)=\frac{\tilde{r}\left(\tilde{r}+2\right)}{k_{u}^{2}+\tilde{r}\left(\tilde{r}+2\right)} \, ,
\ee
which indeed coincides with the Fourier transform of $P_{\text{free},\tilde{r}}\left(x\right)$ from Eq.~\eqref{Pfreer1D}.

All of the moments of the distribution $\mathcal{P}_{\text{free},\tilde{r}}^{\text{st}}\left(\vect{r}\right)$ can be read off Eq.~\eqref{Qfreersol}, since $\mathcal{Q}_{\text{free},\tilde{r}}^{\text{st}}\left(\vect{k}\right)$ is the characteristic function of the distribution. The Taylor-series expansion of $\mathcal{Q}_{\text{free},\tilde{r}}^{\text{st}}\left(k_{u},k_{v}\right)$ around $k_u = k_v = 0$, up to quadratic order in $k_u$ and $k_v$, is
\be
\label{QfreeSeries}
\mathcal{Q}_{\text{free},\tilde{r}}^{\text{st}}\left(k_{u},k_{v}\right)=1-\frac{k_{u}^{2}+k_{v}^{2}}{\tilde{r}\left(\tilde{r}+2\right)}+\frac{\left(6\tilde{r}+8\right)k_{u}^{2}k_{v}^{2}}{\tilde{r}^{2}\left(\tilde{r}+2\right)^{2}\left(\tilde{r}+4\right)} + \dots .
\ee
Of particular interest are the correlation functions that describe the statistical dependence between $u$ and $v$. The first of these is their covariance $\left\langle uv\right\rangle -\left\langle u\right\rangle \left\langle v\right\rangle $ (where angular brackets denote averaging over the steady-state distribution $\mathcal{P}_{\text{free},\tilde{r}}^{\text{st}}$) which vanishes because the coefficients of $k_u k_v$, $k_u$ and $k_v$ in Eq.~\eqref{QfreeSeries} vanish. The lowest nonvanishing correlation function of $x$ and $y$ is therefore
\bea
&&\left\langle u^{2}v^{2}\right\rangle -\left\langle u^{2}\right\rangle \left\langle v^{2}\right\rangle =\left.\partial_{k_{u}}^{2}\partial_{k_{v}}^{2}\mathcal{Q}_{\text{free},\tilde{r}}^{\text{st}}\left(k_{u},k_{v}\right)\right|_{k_{u}=k_{v}=0}\nn\\
&&-\left.\partial_{k_{u}}^{2}\mathcal{Q}_{\text{free},\tilde{r}}^{\text{st}}\left(k_{u},k_{v}\right)\right|_{k_{u}=k_{v}=0}\left.\partial_{k_{v}}^{2}\mathcal{Q}_{\text{free},\tilde{r}}^{\text{st}}\left(k_{u},k_{v}\right)\right|_{k_{u}=k_{v}=0} \nn\\
&&=\frac{4\left(5\tilde{r}+4\right)}{\tilde{r}^{2}\left(\tilde{r}+2\right)^{2}\left(\tilde{r}+4\right)} \, .
%
\eea

 In fact, one can also consider stochastic resetting with an additional confining harmonic potential, with $\mu > 0$. This setting has not been considered even for a 1D RTP. Using a renewal approach that is very similar to the one that gives Eq.~\eqref{PstfreeAndLaplace}, one can show that the steady-state distribution is given by
\be
\label{steadyStateResetting1D}
P_{\tilde{r}}\left(x\right)=\tilde{r}\int_{0}^{\infty}e^{-\tilde{r}\tau}P\left(x,\tau\right)d\tau=\tilde{r}\tilde{P}\left(x,\tilde{r}\right)
\ee
where $P\left(x,\tau\right)$ is the time-dependent position distribution in the absence of resetting, and $\tilde{P}\left(x,\tilde{r}\right)$ is its Laplace transform, which is given above in Eq.~\eqref{Ptildesol}. 
Plugging Eq.~\eqref{Ptildesol} into \eqref{steadyStateResetting1D}, we obtain
\be
\label{steadyStateResetting1Dexplicit}
P_{\tilde{r}}\left(x\right)=\tilde{r}B\left(\tilde{r}\right)z^{\bar{\gamma}+\frac{\tilde{r}}{\mu}-1}\,_{2}F_{1}\left(1-\bar{\gamma},\bar{\gamma};\bar{\gamma}+\frac{\tilde{r}}{\mu};z\right)\,,
\ee
where we recall that $z=\left(1-\mu|x|/v_{0}\right)/2$, and $B(\tilde{r})$ is defined in Eq.~\eqref{Bsdef} above.
$P_{\tilde{r}}\left(x\right)$ is plotted in Fig.~\ref{fig:Prx}. As seen in the figure, the resetting causes the distribution to be localized closer to the center of the trap. In the limit $\tilde{r}\gg1$, the trap becomes unimportant and one recovers the result for a resetting free particle (not shown).
One could consider the 2D case as well, following a similar approach to the one that we used in the free case, i.e., by performing a spatial Fourier transform. However, the calculation would become more technically involved, without presenting much useful additional physics, so we will not present it here.

\subsection{Higher dimensions and/or other geometries}
Our results can immediately be extended to an RTP in 3D, whose orientation vector points towards one of the 8 vertices of a cube centered at the origin, and stochastically jumps between adjacent vertices at a constant rate. 
One simply has to choose a coordinate system $(u,v,w)$ such that the vertices of the cube are in the directions $\pm\hat{u}\pm\hat{v}\pm\hat{w}$.
The problem then decouples into statistically-independent problems for $u$, $v$ and $w$, that each corresponds to a harmonically-confined RTP in 1D.
One can similarly extend to dimensions higher than 3 as well.

\begin{figure}[ht]
\includegraphics[width=0.5\linewidth,clip=]{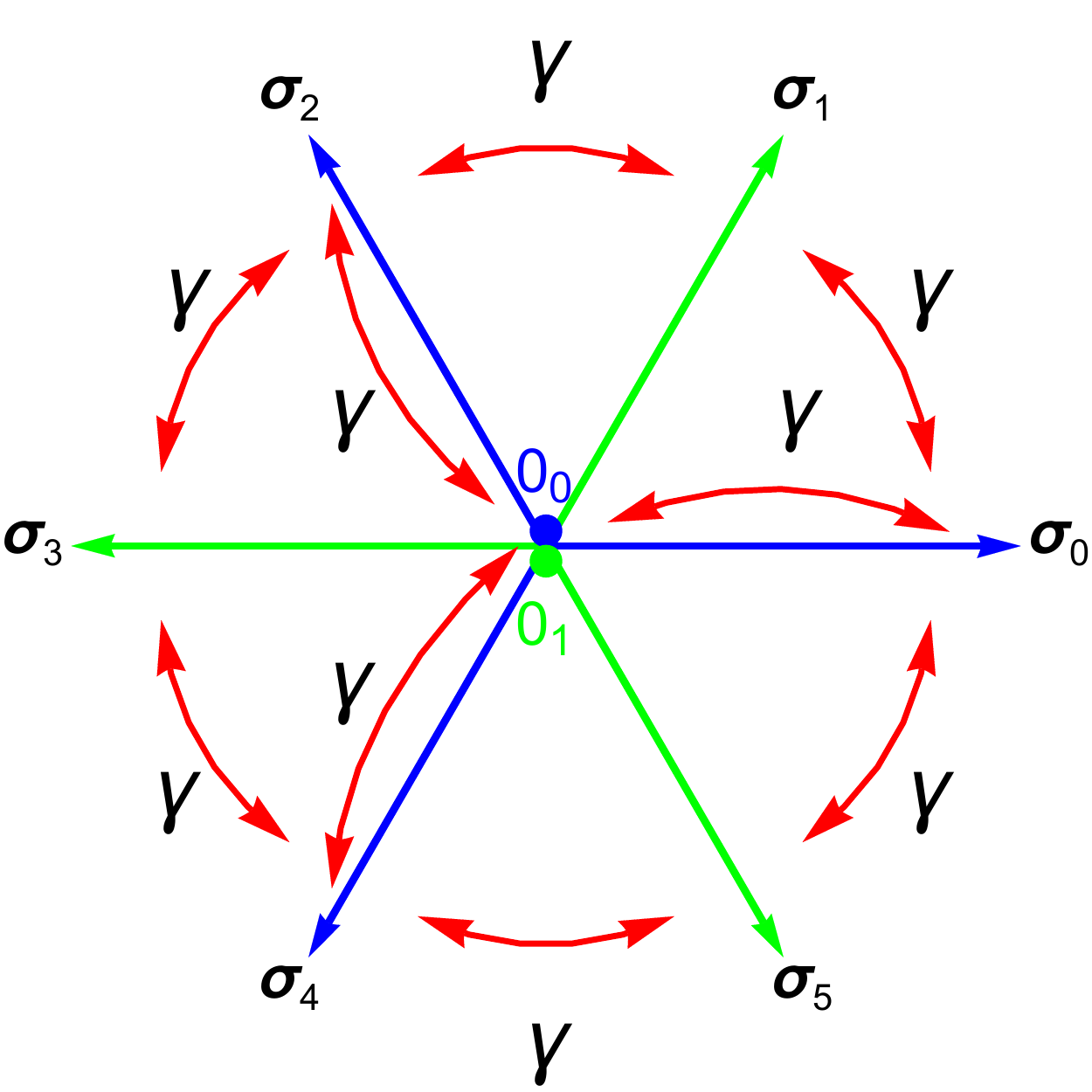}
\caption{A schematic representation of the dynamics of the orientation vector $\vect{\sigma} (t)$ in the 2D RTP hexagonal model described in the text. The possible orientations are the 6 vertices of a regular hexagon $\vect{\sigma}_{0},\dots\vect{\sigma}_{5}$, and two distinct rest states $0_0$ and $0_1$ for which $\vect{\sigma} = 0$.
Transitions occur between adjacent vertices of the hexagon, and between rest states and vertices whose indices have the same parity (the different parities are indicated by different colors in the figure). All transitions occur at the same, constant rate $\gamma$.
For clarity of the figure, the transitions between the rest state $0_1$ and the states $\vect{\sigma}_{1}, \vect{\sigma}_{3}, \vect{\sigma}_{5}$ are not indicated, and the rest states are not placed exactly at the origin.}
\label{fig:Hexagonal}
\end{figure}

Other geometrical configurations for the orientation $\vect{\sigma}$ can also be considered, some of which are also exactly solvable. Let us introduce a planar hexagonal model, in which $\vect{\sigma}$ can take 8 possible values: six vertices of a regular hexagon, 
$\vect{\sigma}_{0},\dots\vect{\sigma}_{5}$ where $\vect{\sigma}_{n} = \left(\cos\frac{n\pi}{3},\sin\frac{n\pi}{3}\right)$,
and two distinct ``rest'' states which we will denote by $0_0$ and $0_1$. Each of the rest states sits at the origin, i.e., $\vect{\sigma} = 0$, and the difference between them corresponds to an internal state of the particle.
The dynamics of the orientation vector are as follows. From a nonzero orientation, the orientation can rotate by 60 degrees, i.e.,
\be
\vect{\sigma}_n \to \vect{\sigma}_{(n\pm1) \, \text{mod} \, 6},
\ee
or else move to the rest state whose index has parity identical to $n$, i.e.,
\be
\vect{\sigma}_n \to 0_{n \, \text{mod}\, 2}.
\ee
From a rest state, the orientation can move to one of 3 nonzero states with the same index parities, as follows:
\be
0_n \to \vect{\sigma}_n , \vect{\sigma}_{n+2}, \vect{\sigma}_{n+4} \, , \quad \text{for} \; n\in\left\{ 0,1\right\} \, .
\ee
Each of the possible transitions in the system occurs at a constant rate which can be taken to be $\gamma$.
The dynamics of $\vect{\sigma}$ are graphically represented in Fig.~\ref{fig:Hexagonal}.

Let us now briefly outline the solution to this hexagonal model.
Remarkably, it turns out that the model can be written as the projection to the $xy$ plane of the 3D ``cube'' model described above, with
\be
\hat{u} \! = \! \sqrt{\frac{2}{3}}\!\left(\!\! \begin{array}{c}
1\\[1mm]
0\\[1mm]
1/\sqrt{2}
\end{array}\!\! \right),\;\hat{v}\! =\! \sqrt{\frac{2}{3}}\!\left(\!\!\begin{array}{c}
\cos\frac{2\pi}{3}\\[1mm]
\sin\frac{2\pi}{3}\\[1mm]
1/\sqrt{2}
\end{array}\!\! \right),\;\hat{w}\! =\! \sqrt{\frac{2}{3}}\!\left(\!\! \begin{array}{c}
\cos\frac{4\pi}{3}\\[1mm]
\sin\frac{4\pi}{3}\\[1mm]
1/\sqrt{2}
\end{array}\!\! \right)\!.\ee
Indeed, one finds that the vertices of the hexagon correspond to the $xy$ projections of 6 of the vertices of the cube, as we now describe. The nonzero orientation states are identified with the vertices of the cube according to
\bea
&\vect{\sigma}_{0}\leftrightarrow\hat{u}-\hat{v}-\hat{w},\quad\vect{\sigma}_{1}\leftrightarrow\hat{u}+\hat{v}-\hat{w},\\
&\vect{\sigma}_{2}\leftrightarrow-\hat{u}+\hat{v}-\hat{w},\quad\vect{\sigma}_{3}\leftrightarrow-\hat{u}+\hat{v}+\hat{w},\\
&\vect{\sigma}_{4}\leftrightarrow-\hat{u}-\hat{v}+\hat{w},\quad\vect{\sigma}_{5}\leftrightarrow\hat{u}-\hat{v}+\hat{w}
\eea
(when considering only the projections into the $xy$ plane of the 3D vectors, and up to a constant of proportionality which can be absorbed into the definition of $v_0$). The two rest states are identified with the two remaining vertices of the cube,
\be
0_{1}\leftrightarrow\hat{u}+\hat{v}+\hat{w},\quad0_{0}\leftrightarrow-\hat{u}-\hat{v}-\hat{w} \, .
\ee

This correspondence can then be immediately exploited in order to solve the hexagonal model. For example, one can take the exact position distribution of the 3D model (which factorizes in the $u,v,w$ coordinates), and then, by marginalizing it along the $z$ direction, one obtains the position distribution of the hexagonal model. We do not present these calculations explicitly here.

Yet another 2D model that can be solved by a decomposition into effective 1D models is that in which the orientation vector takes one of the 9 possible values $\sigma_{x}\hat{x}+\sigma_{y}\hat{y}$, with $\sigma_{x},\sigma_{y}\in\left\{ -1,0,1\right\} $, and where the possible transitions are those for which exactly one of the two components $\sigma_{x},\sigma_{y}$ changes by $\pm 1$. All possible transitions occur at the same rate $\gamma$.
This model decomposes in the $x$ and $y$ coordinates, i.e., the processes $x(t)$ and $y(t)$ are statistically independent, and each of them is described by a 1D RTP model with a noise $\sigma(t)$, changing by $\pm 1$ between the values $\left\{ -1,0,1\right\} $ where all transitions that are possible occur at rate $\gamma$.
%



\section{Discussion}
\label{sec:discussion}

To summarize, we calculated the exact time-dependent distribution $P\left(\vect{r},t\right)$ of the position $\vect{r}$ of an RTP in 2D whose orientation stochastically rotates by 90 degrees, confined by an external harmonic potential. In particular, we found the exact steady-state distribution $P_{\text{st}}\left(\vect{r}\right)$ that is reached in the long-time limit, and also $P\left(\vect{r},t\right)$ for a ``free'' RTP (in the absence of an external potential).
We achieved this by observing that in a properly-chosen coordinate system, the 2D problem decouples into statistically-independent 1D problems, whose solution has been exactly known previously.
We extended these results in several directions. In particular, we showed how to account for diffusion of the RTP,  extended the results to particular RTP models in dimension higher than 2, to two harmonically-interacting RTPs in 2D, and considered stochastic resetting of the RTP's position.
It is worth noting that a decomposition analogous to that of our $P\left(\vect{r},t\right)$ holds for a random walker hopping on a 2D square lattice in discrete time (but not in continuous time). This is a classical result that has been known for quite some time \cite{Spitzer64, Csaki97}.


It would be interesting to try to extend these results to anharmonic potentials and/or to other 2D RTP models, whose orientation changes discontinuously in time, but not by a 90-degree rotation. This presents a major challenge because the equation of motion would then not decouple in the $(u,v)$ coordinates as in the case studied here.
We hope that the theoretical insight that is gained from the exact solution of the particular case studied may shed light on the more general case, perhaps enabling approximate solutions in certain limiting cases.

Another interesting direction for future research is that of systems of many RTPs. 1D chains and gases of RTPs were studied in Refs.~\cite{PBV19, Singh21, ARYL21, Cates22, ARYL22, SSB22}, and it would be interesting to investigate the 2D case.


\bigskip

{\it Acknowledgments:} 
NRS thanks Oded Farago for a collaboration on related topics.
This research was supported by ANR grant ANR-17-CE30-0027-01 RaMaTraF.

\bigskip\bigskip

\begin{widetext} 

\appendix

\section{Proving the Green's function decomposition via the Fokker-Planck approach}
\label{app:Green}

As described in the main text, the decomposition \eqref{GreenDecomposition} of the Green's function follows immediately from the fact that Eqs.~\eqref{udot} and \eqref{vdot} are decoupled, together with the statistical independence of the processes $\sigma_u(t)$ and $\sigma_v(t)$.
Nevertheless, as a useful check, we recover Eq.~\eqref{GreenDecomposition} by using a Fokker-Planck (FP) approach. Although it is rather technical, it may be more natural to some readers, as this approach was used in several previous works.

The joint distribution $\mathcal{P}_i(x,y,t)$ of the position $(x,y)$ and orientation $i = E,N,W,S$ of the RTP evolves according to the FP equation \cite{Basu20, Basu20footnote}
\bea
\frac{\partial}{\partial t} \mathcal{P}_E(x,y,t) &=& \frac{\partial}{\partial x}\left[\left(\mu x-\sqrt{2}\,v_{0}\right)\mathcal{P}_{E}\right]+\frac{\partial}{\partial y}\left(\mu y\mathcal{P}_{E}\right)+\gamma\left(\mathcal{P}_{N}+\mathcal{P}_{S}\right)-2\gamma\mathcal{P}_{E}\,, \label{eq:FP_2d1}\\
\frac{\partial}{\partial t} \mathcal{P}_N(x,y,t) &=&  \frac{\partial}{\partial x}\left(\mu x\mathcal{P}_{N}\right)+\frac{\partial}{\partial y}\left[\left(\mu y-\sqrt{2}\,v_{0}\right)\mathcal{P}_{N}\right]+\gamma\left(\mathcal{P}_{E}+\mathcal{P}_{W}\right)-2\gamma\mathcal{P}_{N}\,, \\
\frac{\partial}{\partial t} \mathcal{P}_W(x,y,t) &=&  \frac{\partial}{\partial x}\left[\left(\mu x+\sqrt{2}\,v_{0}\right)\mathcal{P}_{W}\right]+\frac{\partial}{\partial y}\left(\mu y\mathcal{P}_{W}\right)+\gamma\left(\mathcal{P}_{N}+\mathcal{P}_{S}\right)-2\gamma\mathcal{P}_{W}\,, \\
\frac{\partial}{\partial t} \mathcal{P}_S(x,y,t) &=&  \frac{\partial}{\partial x}\left(\mu x\mathcal{P}_{S}\right)+\frac{\partial}{\partial y}\left[\left(\mu y+\sqrt{2}\,v_{0}\right)\mathcal{P}_{S}\right]+\gamma\left(\mathcal{P}_{E}+\mathcal{P}_{W}\right)-2\gamma\mathcal{P}_{S}\,.\label{eq:FP_2d}
\eea
However, it is far more convenient to solve the problem in the $u,v$ coordinates \eqref{uvdef}. The FP equations that correspond to the Langevin equations \eqref{udot} and \eqref{vdot} are
\bea
\label{eq:FP2DuE}
\frac{\partial}{\partial t}\mathcal{P}_{E}\left(u,v,t\right)&=&\frac{\partial}{\partial u}\left[\left(\mu u-v_{0}\right)\mathcal{P}_{E}\right]+\frac{\partial}{\partial v}\left[\left(\mu v-v_{0}\right)\mathcal{P}_{E}\right]+\gamma\left(\mathcal{P}_{N}+\mathcal{P}_{S}\right)-2\gamma\mathcal{P}_{E}\,,\\
\frac{\partial}{\partial t}\mathcal{P}_{N}\left(u,v,t\right)&=&\frac{\partial}{\partial u}\left[\left(\mu u-v_{0}\right)\mathcal{P}_{N}\right]+\frac{\partial}{\partial v}\left[\left(\mu v+v_{0}\right)\mathcal{P}_{N}\right]+\gamma\left(\mathcal{P}_{E}+\mathcal{P}_{W}\right)-2\gamma\mathcal{P}_{N}\,,\\
\frac{\partial}{\partial t}\mathcal{P}_{W}\left(u,v,t\right)&=&\frac{\partial}{\partial u}\left[\left(\mu u+v_{0}\right)\mathcal{P}_{W}\right]+\frac{\partial}{\partial v}\left[\left(\mu v+v_{0}\right)\mathcal{P}_{W}\right]+\gamma\left(\mathcal{P}_{N}+\mathcal{P}_{S}\right)-2\gamma\mathcal{P}_{W}\,,\\
\label{eq:FP2DuS}
\frac{\partial}{\partial t}\mathcal{P}_{S}\left(u,v,t\right)&=&\frac{\partial}{\partial u}\left[\left(\mu u+v_{0}\right)\mathcal{P}_{S}\right]+\frac{\partial}{\partial v}\left[\left(\mu v-v_{0}\right)\mathcal{P}_{S}\right]+\gamma\left(\mathcal{P}_{E}+\mathcal{P}_{W}\right)-2\gamma\mathcal{P}_{S}\,.
\eea
We now wish to show that the solutions to these equations decompose to solutions of the FP equations for a 1D RTP,
\bea
\label{FP1dplus}
\frac{\partial}{\partial t}P_{+}\left(x,t\right)&=&\frac{\partial}{\partial u}\left[\left(\mu u-v_{0}\right)P_{+}\right]+\gamma P_{-}-\gamma P_{+}\,,\\
\label{FP1dminus}
\frac{\partial}{\partial t}P_{-}\left(x,t\right)&=&\frac{\partial}{\partial u}\left[\left(\mu u+v_{0}\right)P_{-}\right]+\gamma P_{+}-\gamma P_{-}\,.
\eea
Indeed, one can verify directly that given any two solutions $P_{\sigma_{u}}^{\left(1\right)}\left(u,t\right)$ and $P_{\sigma_{v}}^{\left(2\right)}\left(v,t\right)$ to the 1D equations \eqref{FP1dplus} and \eqref{FP1dminus},
\be
\label{PuvtAnsatz}
\mathcal{P}_{\sigma_{u},\sigma_{v}}\left(u,v,t\right)=P_{\sigma_{u}}^{\left(1\right)}\left(u,t\right)P_{\sigma_{v}}^{\left(2\right)}\left(v,t\right)
\ee
is a solution to Eqs.~\eqref{eq:FP2DuE}-\eqref{eq:FP2DuS}, under the identification
\be
E\equiv\left(+,+\right),\quad N\equiv\left(+,-\right),\quad W\equiv\left(-,-\right),\quad S\equiv\left(-,+\right)
\ee
between the possible orientations of $\vect{\sigma}$ and the corresponding signs of its components $\sigma_u$ and $\sigma_v$.
Let us demonstrate that this is indeed the case. Taking a time derivative of Eq.~\eqref{PuvtAnsatz} for $\sigma_u = \sigma_v = +$, one obtains, using Eq.~\eqref{FP1dplus},
\bea
\frac{\partial}{\partial t}\mathcal{P}_{E}\left(u,v,t\right)&=&\frac{\partial}{\partial t}P_{+}^{\left(1\right)}\left(u,t\right)P_{+}^{\left(2\right)}\left(v,t\right)+P_{+}^{\left(1\right)}\left(u,t\right)\frac{\partial}{\partial t}P_{+}^{\left(2\right)}\left(v,t\right) \nn\\
&=&\left\{ \frac{\partial}{\partial u}\left[\left(\mu u-v_{0}\right)P_{+}^{\left(1\right)}\left(u,t\right)\right]+\gamma P_{-}^{\left(1\right)}\left(u,t\right)-\gamma P_{+}^{\left(1\right)}\left(u,t\right)\right\} P_{+}^{\left(2\right)}\left(v,t\right) \nn\\
&+& P_{+}^{\left(1\right)}\left(u,t\right)\left\{ \frac{\partial}{\partial v}\left[\left(\mu v-v_{0}\right)P_{+}^{\left(2\right)}\left(v,t\right)\right]+\gamma P_{-}^{\left(2\right)}\left(v,t\right)-\gamma P_{+}^{\left(2\right)}\left(v,t\right)\right\} \nn\\
&=&\frac{\partial}{\partial u}\left[\left(\mu u-v_{0}\right)\mathcal{P}_{E}\right]+\gamma\mathcal{P}_{S}-\gamma\mathcal{P}_{E}+\frac{\partial}{\partial v}\left[\left(\mu v-v_{0}\right)\mathcal{P}_{E}\right]+\gamma\mathcal{P}_{N}-\gamma\mathcal{P}_{E} \,,
\eea
which indeed coincides with the right hand side of Eq.~\eqref{eq:FP2DuE}.

The decomposition \eqref{GreenDecomposition} of the Green's function given in the main text is a particular case of Eq.~\eqref{PuvtAnsatz} in which
\be
P_{\sigma_{u}}^{\left(1\right)}\left(u,t=t'\right)=\delta\left(u-u'\right)\delta_{\sigma_{u},\sigma'_{u}}
\ee
and similarly in the $v$ direction. Similarly, the decomposition \eqref{Pxyt} of the position distribution is a particular case of Eq.~\eqref{PuvtAnsatz} in which the initial condition is 
\be
P_{\sigma_{u}}^{\left(1\right)}\left(u,t=0\right)=\frac{1}{2}\delta\left(u\right)
\ee
and similarly in the $v$ direction.

\section{Marginal distribution of $x$}
\label{app:marginal}

In \cite{Basu20}, the marginal steady-state distribution of $x$ was calculated for the 2D RTP.
As a useful check of \eqref{Pstsol} of the main text, we can reproduce this result. By using
$x=\left(u+v\right)/\sqrt{2}$,
the marginal distribution that we predict is
\bea
\label{marginal1}
P_{\text{marginal,st}}\left(x\right)&=&\int_{-\infty}^{\infty}du\int_{-\infty}^{\infty}dv\mathcal{P}_{\text{st}}\left(u,v\right)\delta\left(x-\frac{u+v}{\sqrt{2}}\right)\nn\\
&=& \sqrt{2}\int_{-\infty}^{\infty}du\mathcal{P}_{\text{st}}\left(u,v=\sqrt{2}x-u\right) \nn\\
&=&\frac{\sqrt{2}}{Z}\int_{\sqrt{2}\,\left|x\right|-v_{0}/\mu}^{v_{0}/\mu}du\left[\left(1-\left(\frac{\mu u}{v_{0}}\right)^{2}\right)\left(1-\left(\frac{\mu}{v_{0}}\right)^{2}\left(\sqrt{2}\,\left|x\right|-u\right)^{2}\right)\right]^{\beta-1},
\eea
where
\be
Z^{-1}=\frac{4\mu^{2}}{2^{4\beta}\left[B\left(\beta,\beta\right)v_{0}\right]^{2}}
\ee
is a normalization factor, and we used the mirror symmetry $P_{\text{marginal,st}}\left(-x\right)=P_{\text{marginal,st}}\left(x\right)$.
The integral in \eqref{marginal1} is in general not so easy to calculate. However, for $\beta = 1$ and $\beta = 2$ it evaluates to
\be
\left.P_{\text{marginal,st}}\left(x\right)\right|_{\beta=1}=\frac{\mu}{2v_{0}}\left(\sqrt{2}-\frac{\mu\left|x\right|}{v_{0}}\right)
\ee
and
\be
\left.P_{\text{marginal,st}}\left(x\right)\right|_{\beta=2}=\frac{9\mu}{8\sqrt{2}v_{0}}\left[\frac{16}{15}-\frac{8}{3}\left(\frac{\mu x}{v_{0}}\right)^{2}+\frac{4\sqrt{2}}{3}\left|\frac{\mu x}{v_{0}}\right|^{3}-\frac{2}{15}\sqrt{2}\left|\frac{\mu x}{v_{0}}\right|^{5}\right]
\ee
respectively, in perfect agreement with \cite{Basu20, Basu20footnote} (note that in \cite{Basu20} an explicit expression for the marginal distribution, in terms of hypergeometric functions, was obtained for arbitrary $\beta > 0$).

We can preform the same check for the time-dependent marginal distribution of the $x$ coordinate of a free RTP, comparing with the exact result from Ref.~\cite{SBS20}.
It turns out to be much simpler to perform the comparison in Fourier space. For simplicity, let us choose units in which $\gamma = v_0 = 1$. The Fourier transform of the distribution \eqref{p1xt.1} of the position of a free 1D RTP is given by \cite{MalakarEtAl18} Eq.~\eqref{Pfreek} of the main text, which we give here again for convenience:
\be
\label{PfreekA}
Q_{\text{free}}\left(k,t\right)=\int_{-\infty}^{\infty}e^{ikx}P_{\text{free}}\left(x,t\right)dx=e^{-t}\left[\cosh\left(t\sqrt{1-k^{2}}\right)+\frac{1}{\sqrt{1-k^{2}}}\sinh\left(t\sqrt{1-k^{2}}\right)\right].
\ee
Now, since $x=\left(u+v\right)/\sqrt{2}$, $u(t)$ and $v(t)$ both being independent and described by the same distribution \eqref{p1xt.1}, the marginal distribution $P_{\text{marginal,free}}\left(x,t\right)$ is given by the convolution of the distributions of $u$ and $v$. In Fourier space, the convolution becomes a product, which (taking into account the factor $\sqrt{2}$) leads to
\be
Q_{\text{marginal,free}}\left(k,t\right)=\int_{-\infty}^{\infty}e^{ikx}P_{\text{marginal,free}}\left(x,t\right)dx=\left[Q_{\text{free}}\left(\frac{k}{\sqrt{2}},t\right)\right]^{2} \, .
\ee
Using Eq.~\eqref{PfreekA}, we find
\be
\label{Pfreeksq}
\left[Q_{\text{free}}\left(k,t\right)\right]^{2} = \frac{e^{-2t}}{2\left(1-k^{2}\right)}\left[-k^{2}+\left(2-k^{2}\right)\cosh\left(2t\sqrt{1-k^{2}}\right)+2\sqrt{1-k^{2}}\sinh\left(2t\sqrt{1-k^{2}}\right)\right] \, ,
\ee
where we used the standard identities
\be
\cosh x\sinh x=\frac{\sinh\left(2x\right)}{2},\quad\cosh^{2}x=\frac{\cosh\left(2x\right)+1}{2},\quad \cosh^{2}x-\sinh^{2}x=1 \, .
\ee
Eq.~\eqref{Pfreeksq} is in perfect agreement with the result of \cite{SBS20, Basu20footnote, SBS20minus}.
In fact, in \cite{SBS20}, the Fourier transform was inverted and an expression for $P_{\text{marginal,free}}\left(x,t\right)$ was obtained.

\end{widetext}

{}


\begin{thebibliography}{}
\bibitem{Basu20footnote} Note that in Refs.~\cite{SBS20, Basu20}, the transition rates are denoted by $\gamma / 2$, and the particle's velocity is $v_0$, whereas in our units, they are given by $\gamma$ and $\sqrt{2} \, v_0$, respectively.
\bibitem{SBS20minus} Our Eq.~\eqref{Pfreeksq} agrees with Eq. (58) in \cite{SBS20} only up to an overall minus sign. This appears to be due to a typo in \cite{SBS20}, because when plugging $k=0$ in Eq.~\eqref{Pfreeksq} we get $Q_{\text{marginal,free}}\left(k=0,t\right) = 1$ as required from the normalization, (whereas the same check gives $-1$ in the corresponding formula from \cite{SBS20}).
\end{thebibliography}
\end{document}